\documentclass[journal]{IEEEtran}
\usepackage{cite}
\usepackage{graphicx}
\usepackage{url}
\usepackage{amsmath}
\usepackage{amssymb}
\usepackage[caption = false, font=footnotesize]{subfig}
\usepackage{float}
\usepackage{booktabs} 

\usepackage{array}
\usepackage{color}
\usepackage{tabularx}
\usepackage{longtable}
\usepackage{tabu}
\usepackage{multirow}
\newcolumntype{L}[1]{>{\raggedright\let\newline\\\arraybackslash\hspace{0pt}}m{#1}}
\newcolumntype{C}[1]{>{\centering\let\newline\\\arraybackslash\hspace{0pt}}m{#1}}
\newcolumntype{R}[1]{>{\raggedleft\let\newline\\\arraybackslash\hspace{0pt}}m{#1}}

\newcommand{\eg}{\textit{e}.\textit{g}.}

\ifCLASSINFOpdf
\else
\fi
\begin{document}
%
\title{Blind Image Quality Assessment Using \\ A Deep Bilinear Convolutional Neural Network}
%
%
%

\author{Weixia~Zhang,
        Kede~Ma,~\IEEEmembership{Member,~IEEE,}
        Jia~Yan,
        Dexiang~Deng,
        and~Zhou~Wang,~\IEEEmembership{Fellow,~IEEE}
\thanks{This work was supported in part by the National Natural Science Foundation of China under Grant 61701351.}
\thanks{Weixia Zhang, Jia Yan, and Dexiang Deng are with the Electronic Information School, Wuhan University, Wuhan, China (e-mail: zhangweixia@whu.edu.cn; yanjia2011@gmail.com; ddx@whu.edu.cn).}
\thanks{Kede Ma is with the Center for Neural Science, New York University, New York, NY 10003, USA (e-mail: kede.ma@nyu.edu).}
\thanks{Zhou Wang is with the Department of Electrical and Computer Engineering, University of Waterloo, Waterloo, ON N2L 3G1, Canada (e-mail: zhou.wang@uwaterloo.ca).}
\thanks{\textcircled{c} 2019 IEEE. Personal use of this material is permitted. Permission from IEEE must be obtained for all other uses, in any current or future media, including reprinting/republishing this material for advertising or promotional purposes, creating new collective works, for resale or redistribution to servers or lists, or reuse of any copyrighted component of this work in other works.}
}

\maketitle

\begin{abstract}
We propose a deep bilinear model for blind image quality assessment (BIQA) that handles both synthetic and authentic distortions. Our model consists of two convolutional neural networks (CNN), each of which  specializes in one distortion scenario. For synthetic distortions, we pre-train a CNN to classify image distortion type and level, where we enjoy large-scale training data. For authentic distortions, we adopt a pre-trained CNN for image classification. The features from the two CNNs are  pooled bilinearly into a unified representation for final quality prediction. We then fine-tune the entire model on target subject-rated databases using a variant of stochastic gradient descent. Extensive experiments demonstrate that the proposed model achieves superior performance on both synthetic and authentic databases. Furthermore, we verify the generalizability of our method on the Waterloo Exploration Database  using the group maximum differentiation competition.
\end{abstract}


%
\IEEEpeerreviewmaketitle
\section{Introduction}
\IEEEPARstart{N}{owadays}, digital images are captured via various mobile cameras, compressed by conventional and advanced techniques~\cite{bovik2010handbook,BalleLS16a}, transmitted through diverse communication channels~\cite{duanmu2017quality}, and stored on different devices. Each stage in the image processing pipeline could introduce unexpected distortions, leading to perceptual quality degradation. Therefore, image quality assessment (IQA) is of great importance to monitoring the quality of images and ensuring the reliability of image processing systems.
It is essential to design accurate and efficient computational models to push IQA from laboratory research to real-world applications~\cite{wang2006modern,rehman2015display}. Among all computational models, we are interested in no-reference or blind IQA (BIQA) methods~\cite{wang2011reduced} because the reference information is often unavailable (or may not exist) in many practical applications.


Previous knowledge-driven BIQA models typically adopt low-level features either hand-crafted~\cite{mittal2012no} or learned~\cite{ye2012unsupervised} to characterize the level of deviations from statistical regularities of natural scenes. Until recently, there has been limited effort towards end-to-end optimized BIQA using deep convolutional neural networks (CNN)~\cite{kang2014convolutional,Ma2018End}, primarily due to the lack of sufficient ground truths such as the mean opinion scores (MOS) for training. A straightforward approach is to fine-tune a CNN pre-trained on ImageNet~\cite{deng2009imagenet} for quality prediction~\cite{kim2017deep}.
The resulting model performs reasonably on the LIVE Challenge Database~\cite{ghadiyaram2016massive} (with authentic distortions), but does not stand out on the LIVE~\cite{sheikh2006statistical} and TID2013~\cite{ponomarenko2013color} databases (with synthetic distortions). Another common strategy is patch-based training, where the patch-level ground truths are either inherited from image-level annotations~\cite{kang2014convolutional} or approximated by full-reference IQA models~\cite{kim2017fully}. This strategy is very effective at learning CNN-based models for synthetic distortions, but fails to handle authentic distortions due to the non-homogeneity of distortions and the absence of reference images. Other methods~\cite{Kang2015Simultaneous,Ma2018End} take advantage of  synthetic degradation processes (\eg, distortion types) to find reasonable initializations for CNN-based models, but cannot be applied to authentic distortions either.

\begin{figure*}[th]
    \centering
    \captionsetup{justification=centering}
    \subfloat[]{\includegraphics[width=0.32\textwidth]{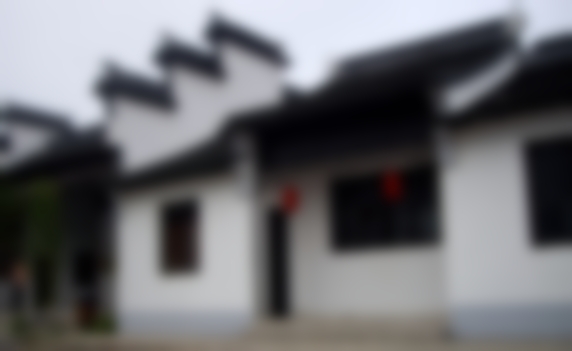}}\hskip.2em
    \subfloat[]{\includegraphics[width=0.32\textwidth]{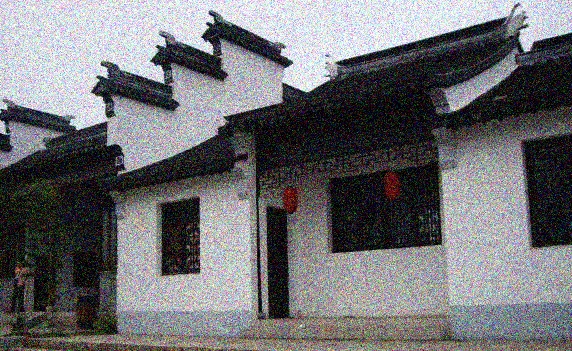}}\hskip.2em
    \subfloat[]{\includegraphics[width=0.32\textwidth]{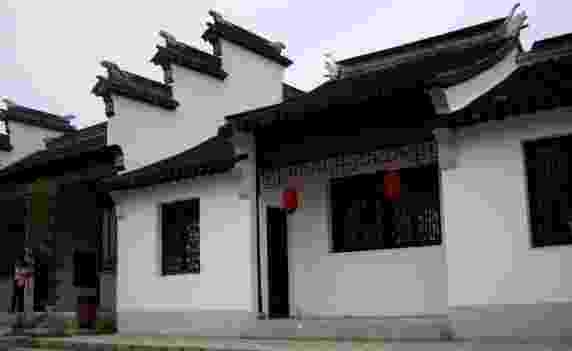}}
    \vspace{2pt}
    \subfloat[]{\includegraphics[width=0.32\textwidth]{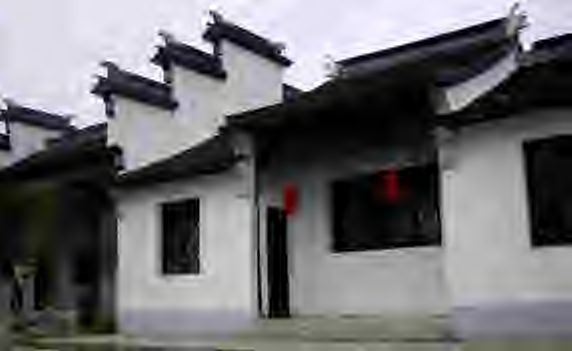}}\hskip.2em
    \subfloat[]{\includegraphics[width=0.32\textwidth]{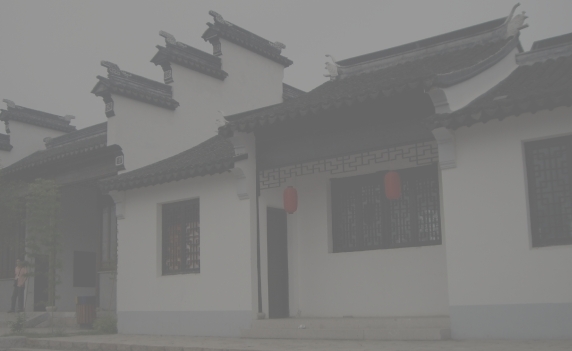}}\hskip.2em
    \subfloat[]{\includegraphics[width=0.32\textwidth]{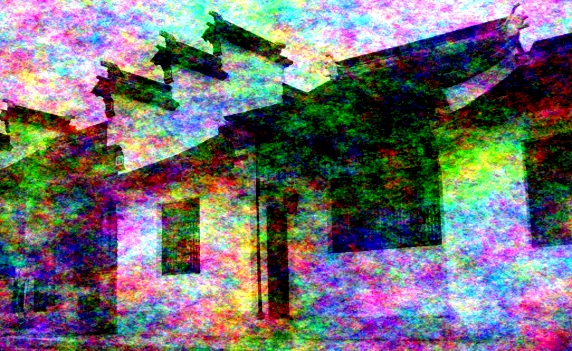}}
    \vspace{2pt}
    \subfloat[]{\includegraphics[width=0.32\textwidth]{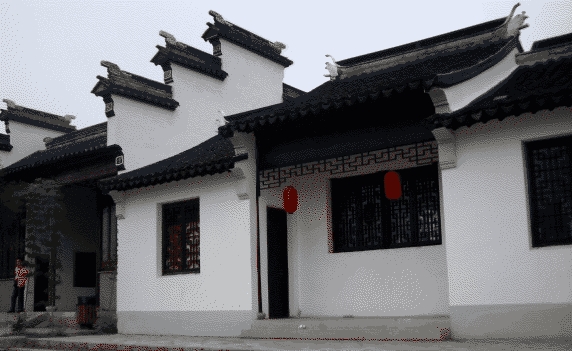}}\hskip.2em
    \subfloat[]{\includegraphics[width=0.32\textwidth]{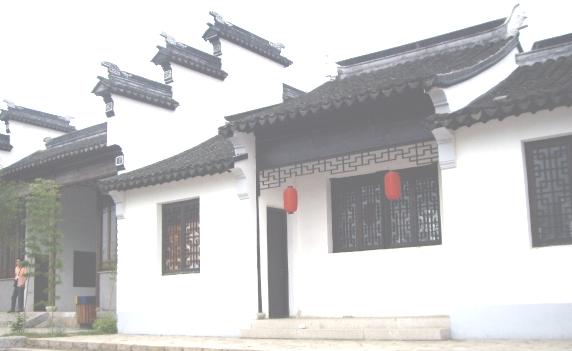}}\hskip.2em
    \subfloat[]{\includegraphics[width=0.32\textwidth]{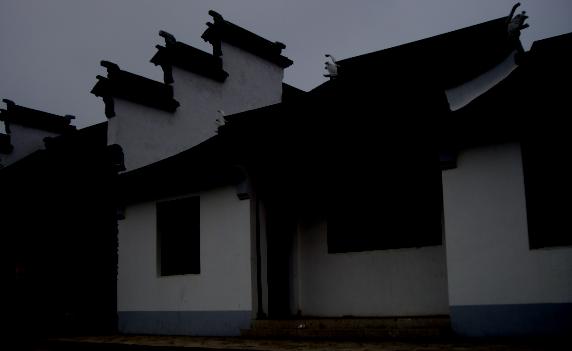}}
    \vspace{2pt}
    \caption{Sample distorted images synthesized from a reference image in the Waterloo Exploration Database~\cite{ma2017waterloo}. (a) Gaussian blur. (b) White Gaussian noise. (c) JPEG compression. (d) JPEG2000 compression. (e) Contrast stretching.  (f) Pink noise. (g) Image color quantization with dithering. (h) Over-exposure. (i) Under-exposure.}\label{fig:distortions}
\end{figure*}

In this work, we aim for an end-to-end solution to BIQA that handles both synthetic and authentic distortions. We first learn two feature sets for the two distortion scenarios separately. For synthetic distortions, inspired by previous studies~\cite{Kang2015Simultaneous,Ma2018End}, we construct a large-scale pre-training set based on the Waterloo Exploration Database~\cite{ma2017waterloo} and the PASCAL VOC Database~\cite{everingham2010pascal}, where the images are synthesized with nine distortion types and two to five distortion levels. We take advantage of known distortion type and level information in the dataset and pre-train a CNN through a multi-class classification task. For authentic distortions, it is difficult to simulate the degradation processes due to their complexities~\cite{ghadiyaram2017perceptual}. Therefore, we opt for another CNN (VGG-16~\cite{simonyan2014very}) pre-trained on ImageNet~\cite{deng2009imagenet} that contains many realistic natural images of different perceptual quality.   We model synthetic and authentic distortions as two-factor variations, and pool the two feature sets bilinearly~\cite{lin2015bilinear} into a unified representation for final quality prediction. The resulting deep bilinear CNN (DB-CNN) is fine-tuned on target subject-rated databases using a variant of the stochastic gradient descent method. Extensive experimental results on five  IQA databases demonstrate the effectiveness of DB-CNN for both synthetic and authentic distortions. Furthermore, through the group MAximum Differentiation (gMAD) competition~\cite{ma2016group}, we find that DB-CNN is more robust than the most recent CNN-based BIQA models~\cite{bosse2016deep,Ma2018End}.


\begin{figure*}[th]
    \centering
    \captionsetup{justification=centering}
    \subfloat[]{\includegraphics[width=0.19\textwidth]{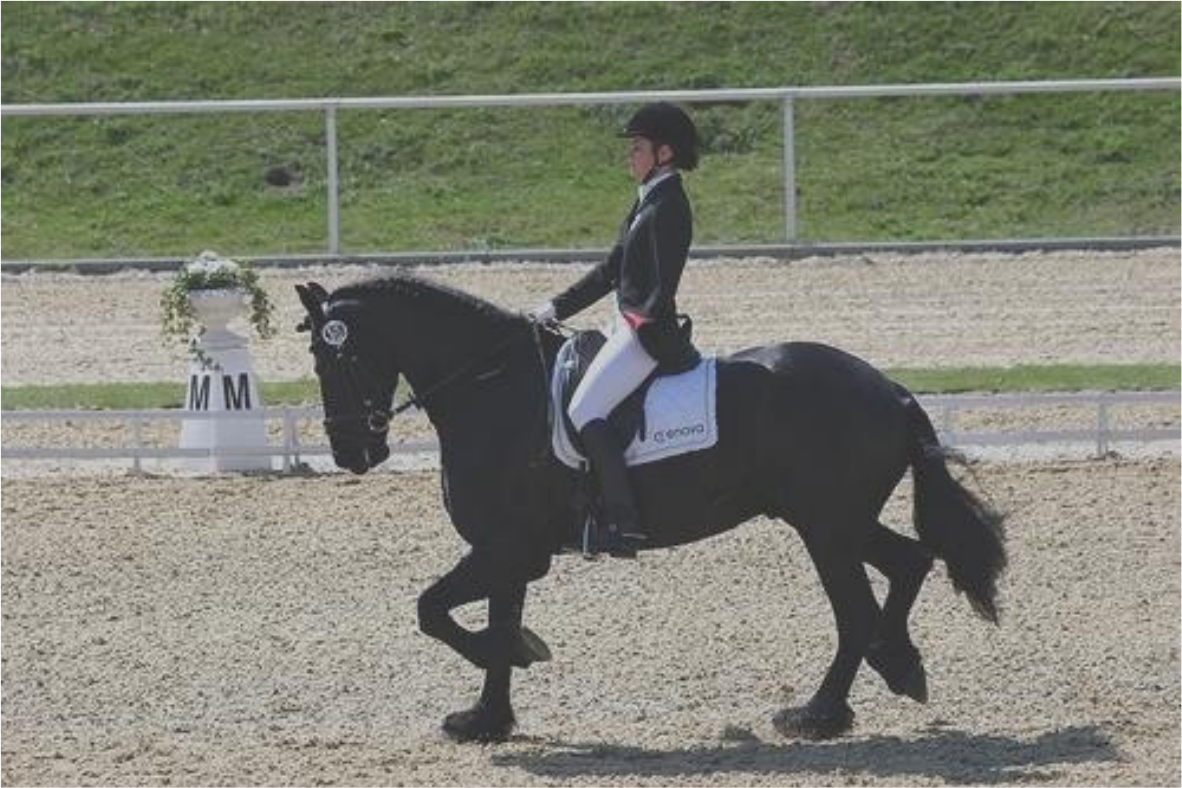}}\hskip.2em
    \subfloat[]{\includegraphics[width=0.19\textwidth]{figs/l1}}\hskip.2em
    \subfloat[]{\includegraphics[width=0.19\textwidth]{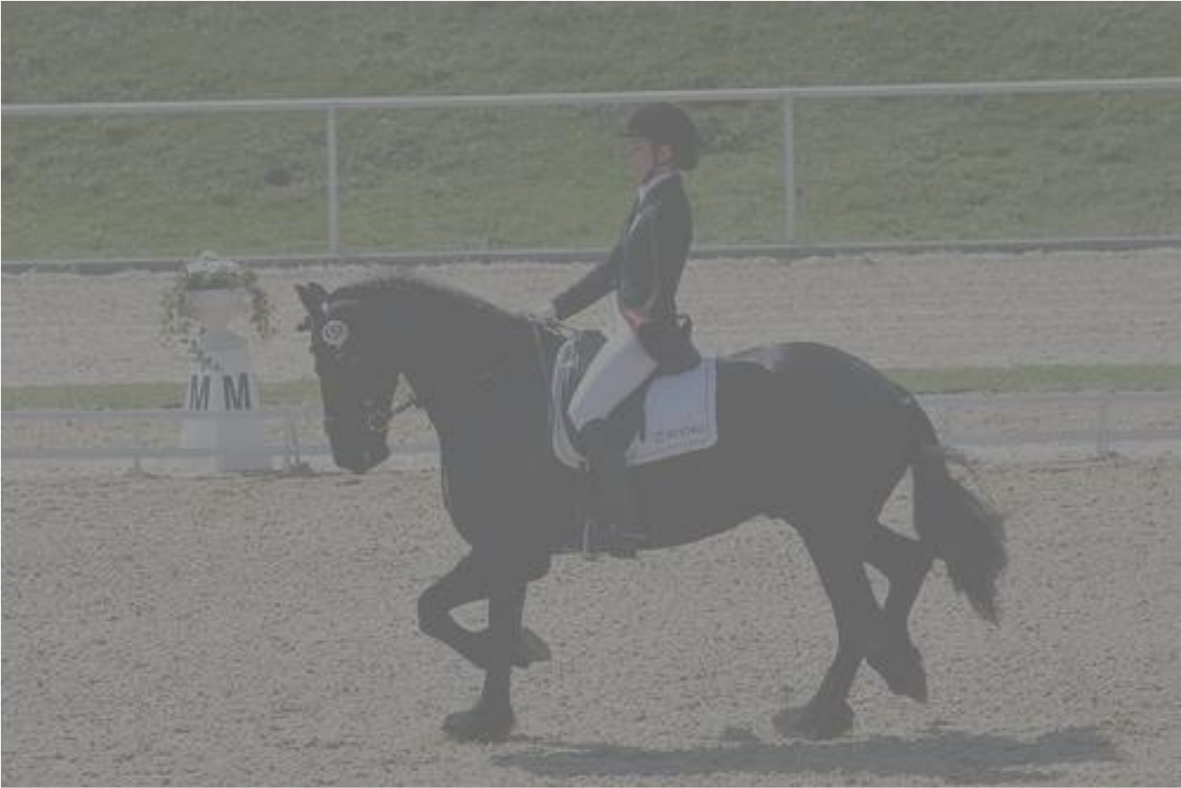}}\hskip.2em
    \subfloat[]{\includegraphics[width=0.19\textwidth]{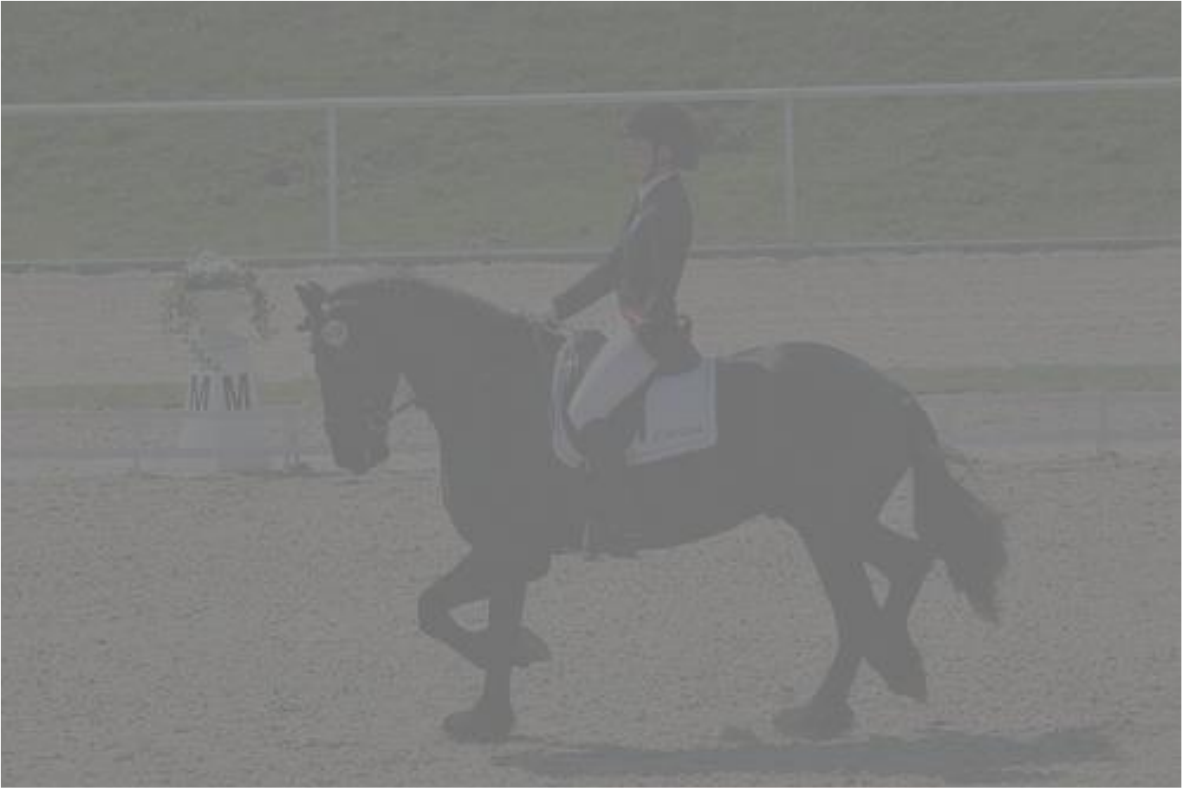}}\hskip.2em
    \subfloat[]{\includegraphics[width=0.19\textwidth]{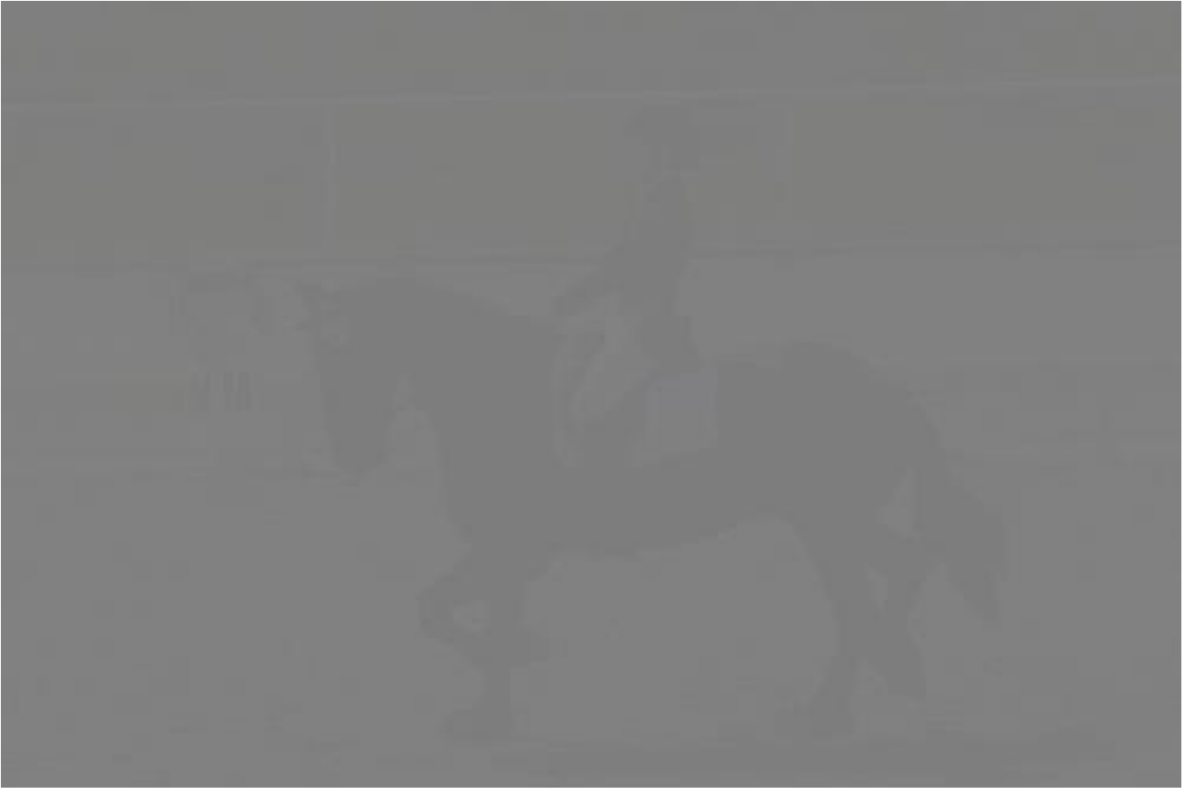}}
    \vspace{2pt}
    \subfloat[]{\includegraphics[width=0.19\textwidth]{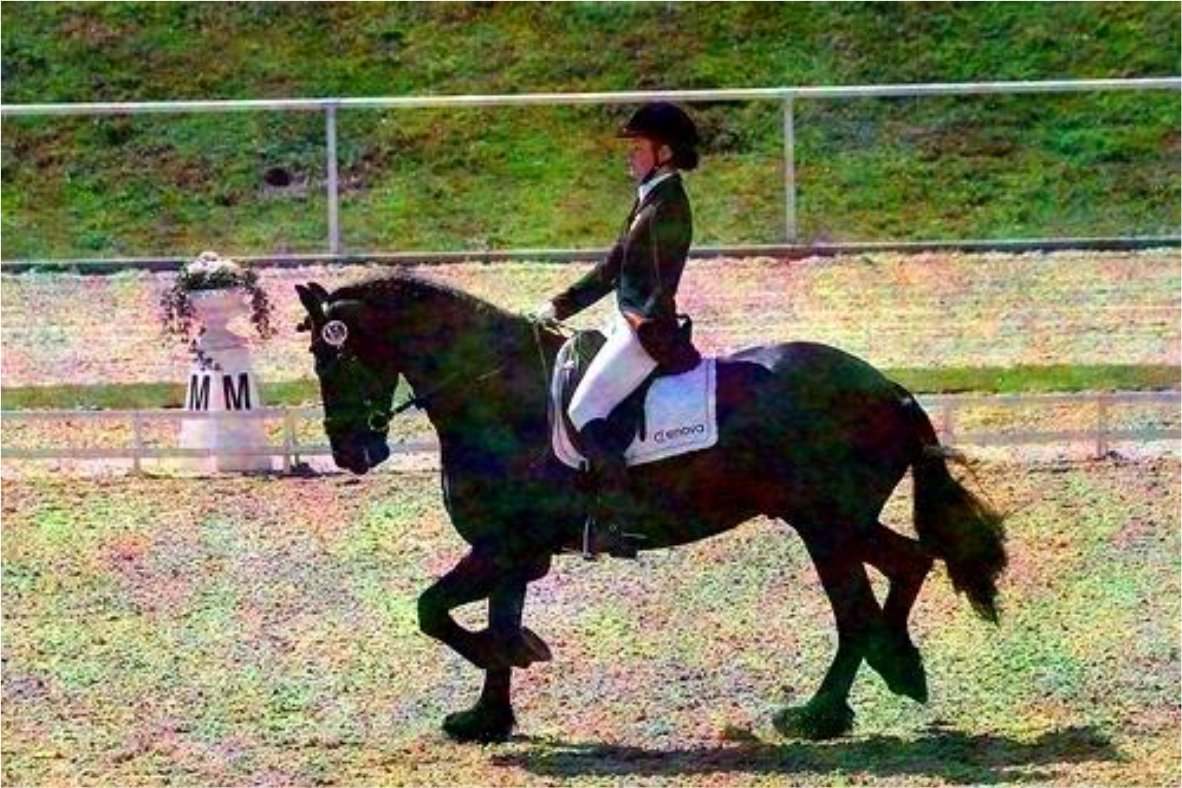}}\hskip.2em
    \subfloat[]{\includegraphics[width=0.19\textwidth]{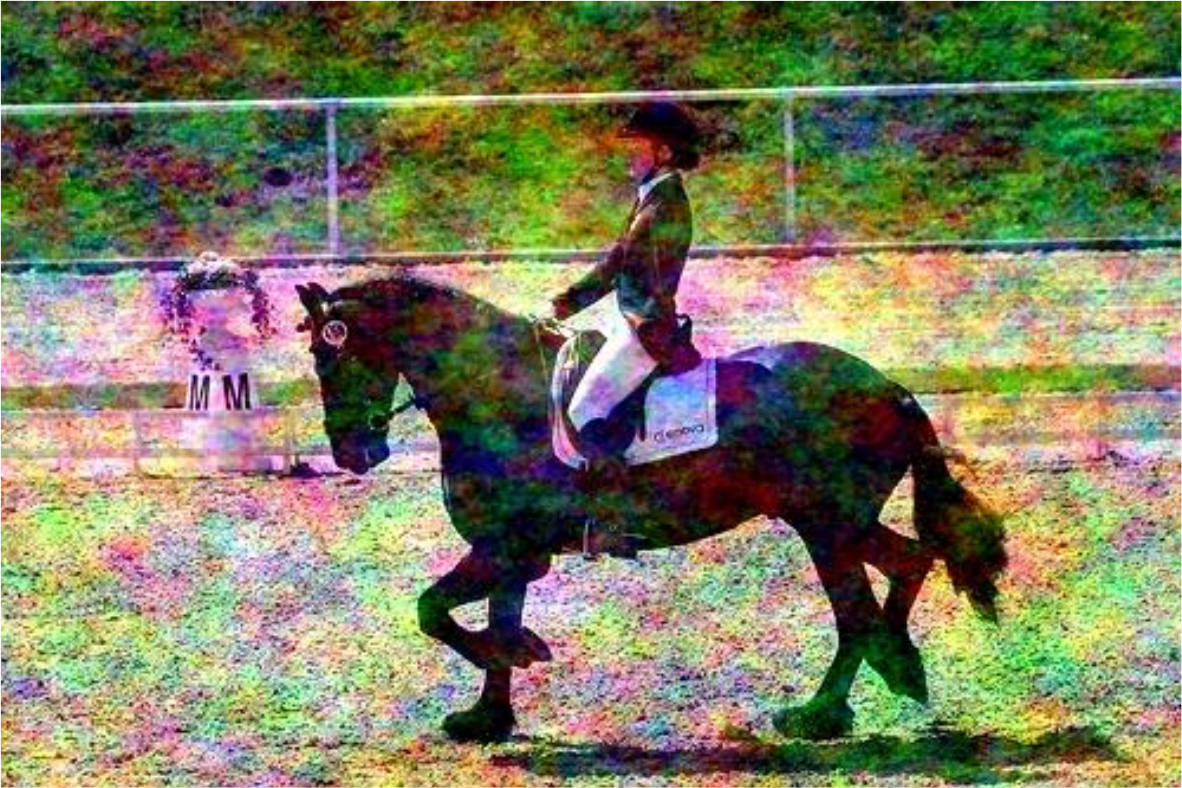}}\hskip.2em
    \subfloat[]{\includegraphics[width=0.19\textwidth]{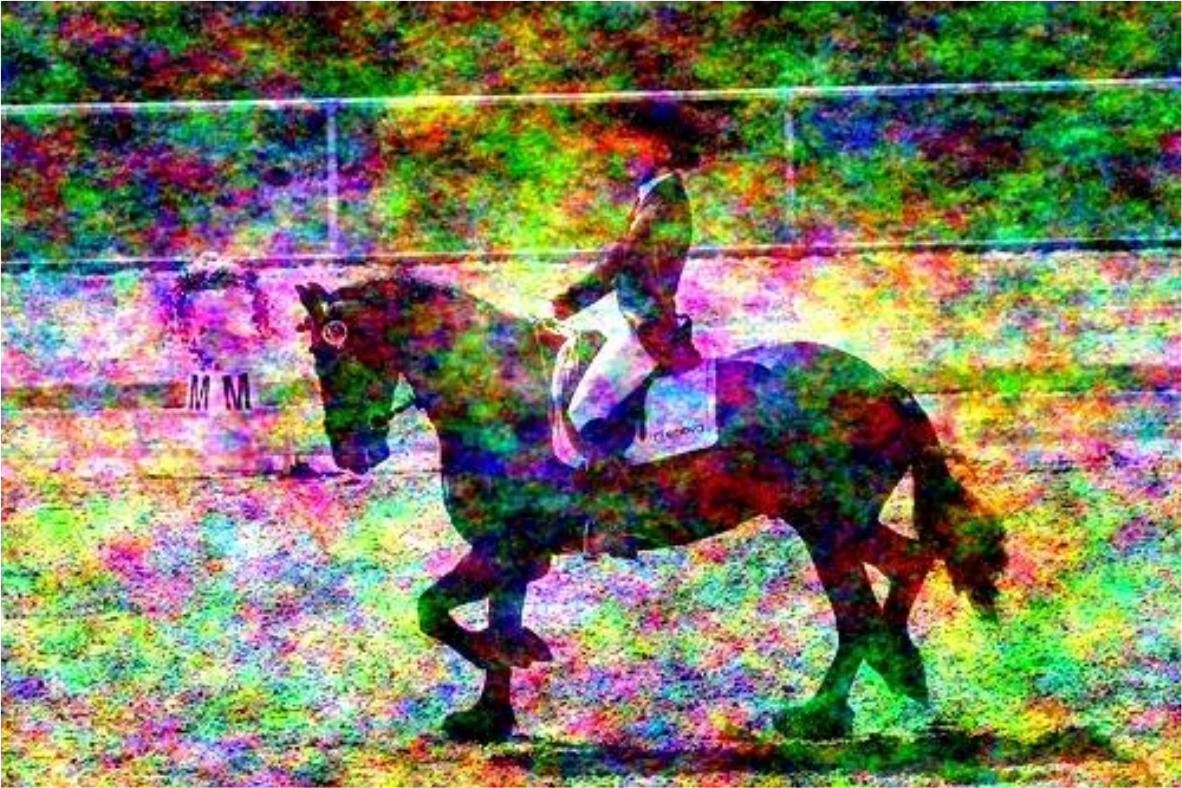}}\hskip.2em
    \subfloat[]{\includegraphics[width=0.19\textwidth]{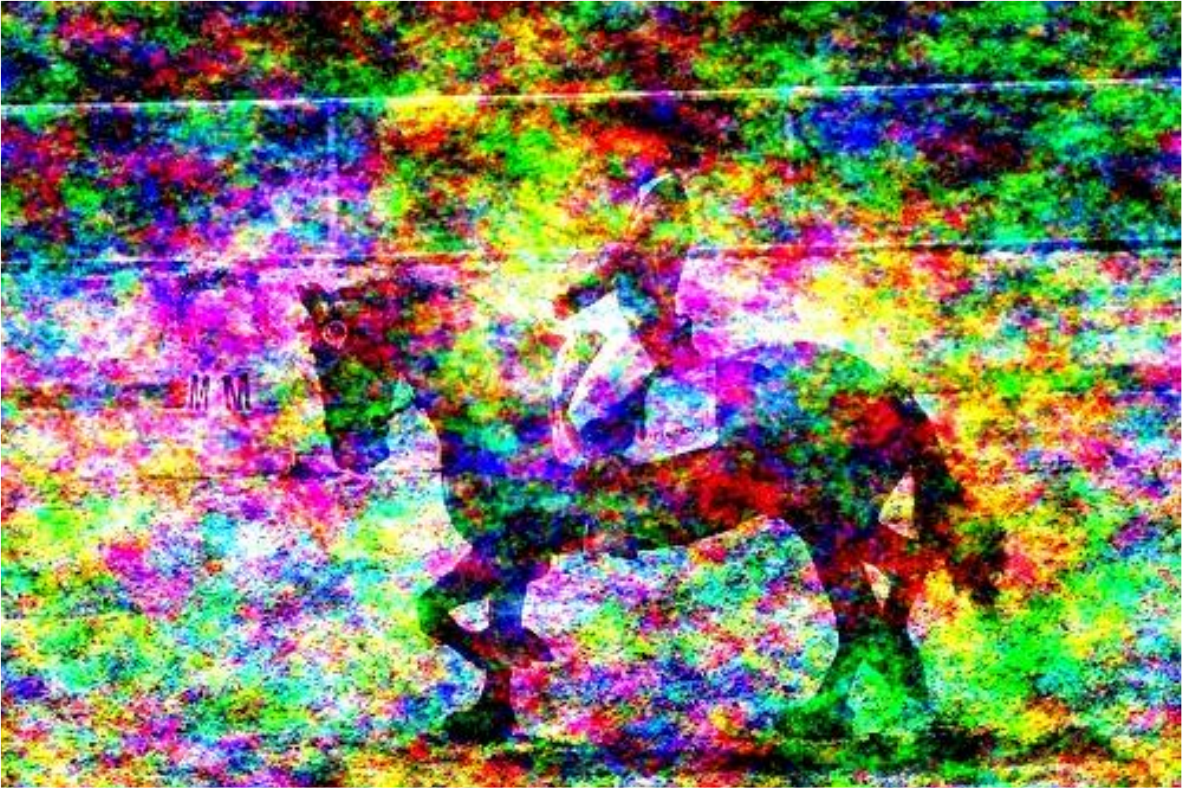}}\hskip.2em
    \subfloat[]{\includegraphics[width=0.19\textwidth]{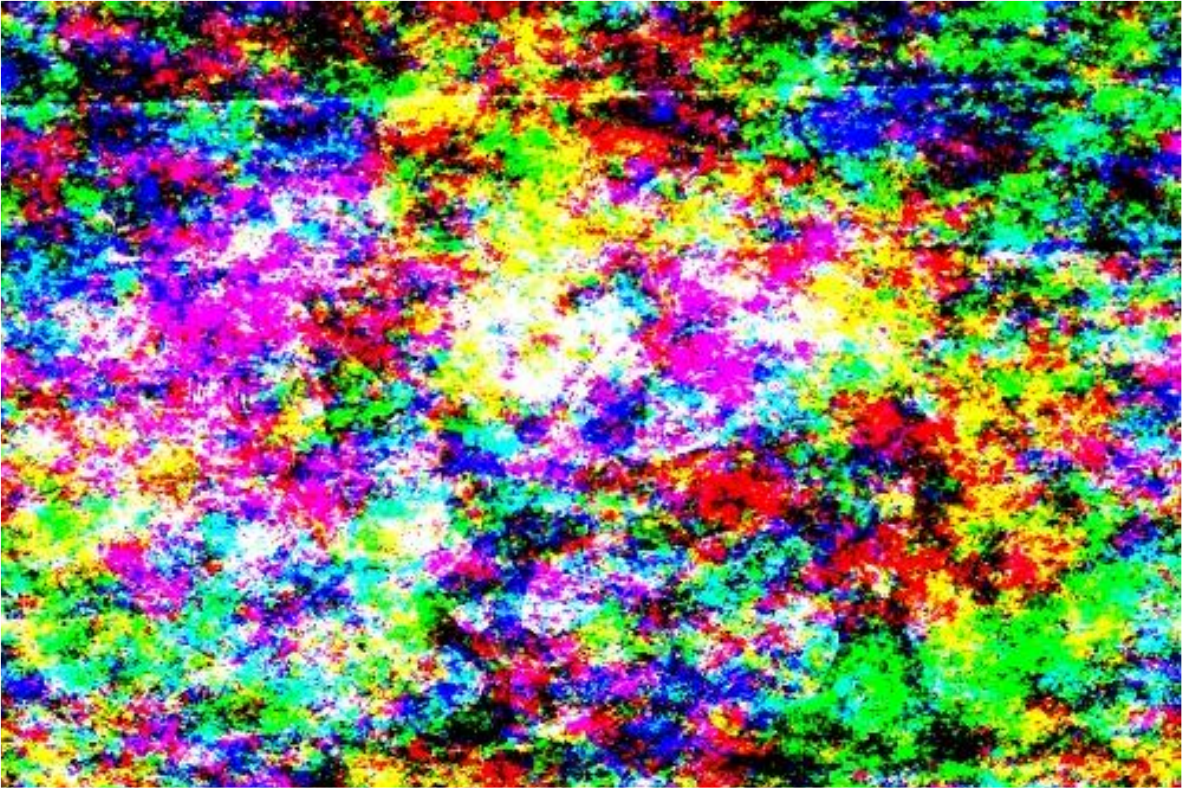}}
    \vspace{2pt}
    \subfloat[]{\includegraphics[width=0.19\textwidth]{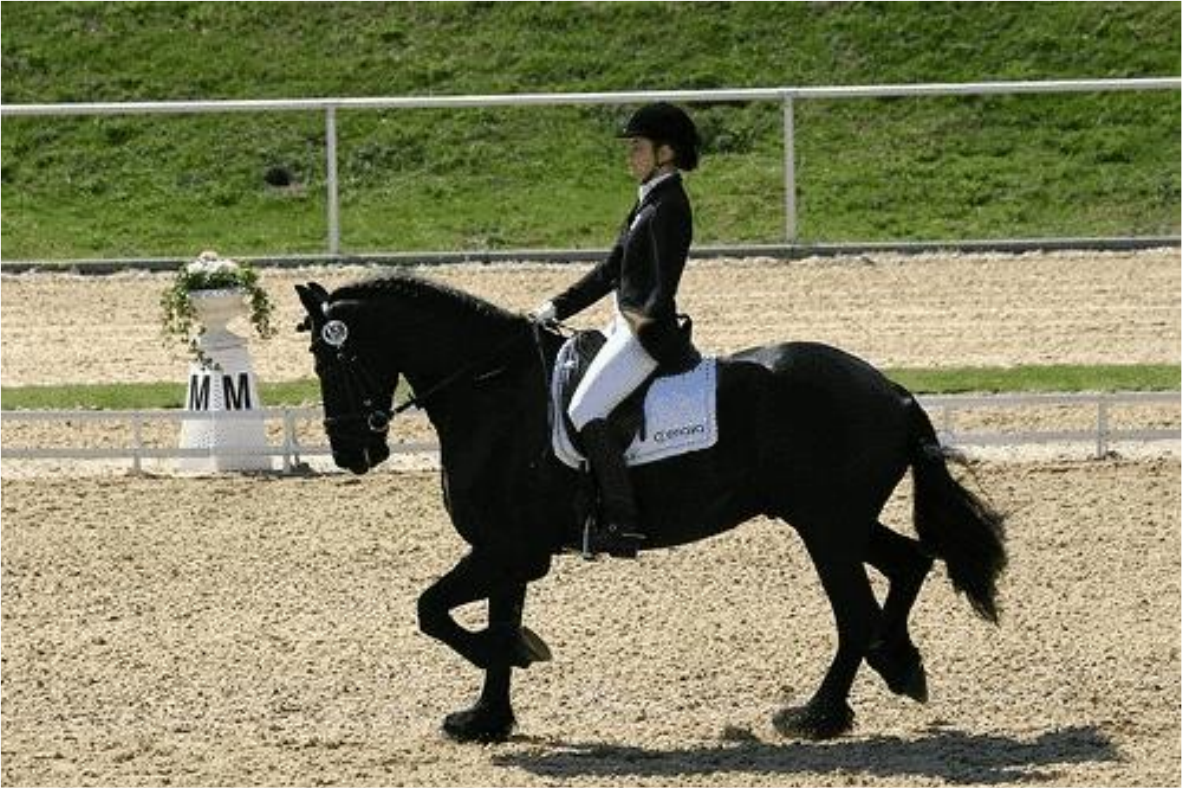}}\hskip.2em
    \subfloat[]{\includegraphics[width=0.19\textwidth]{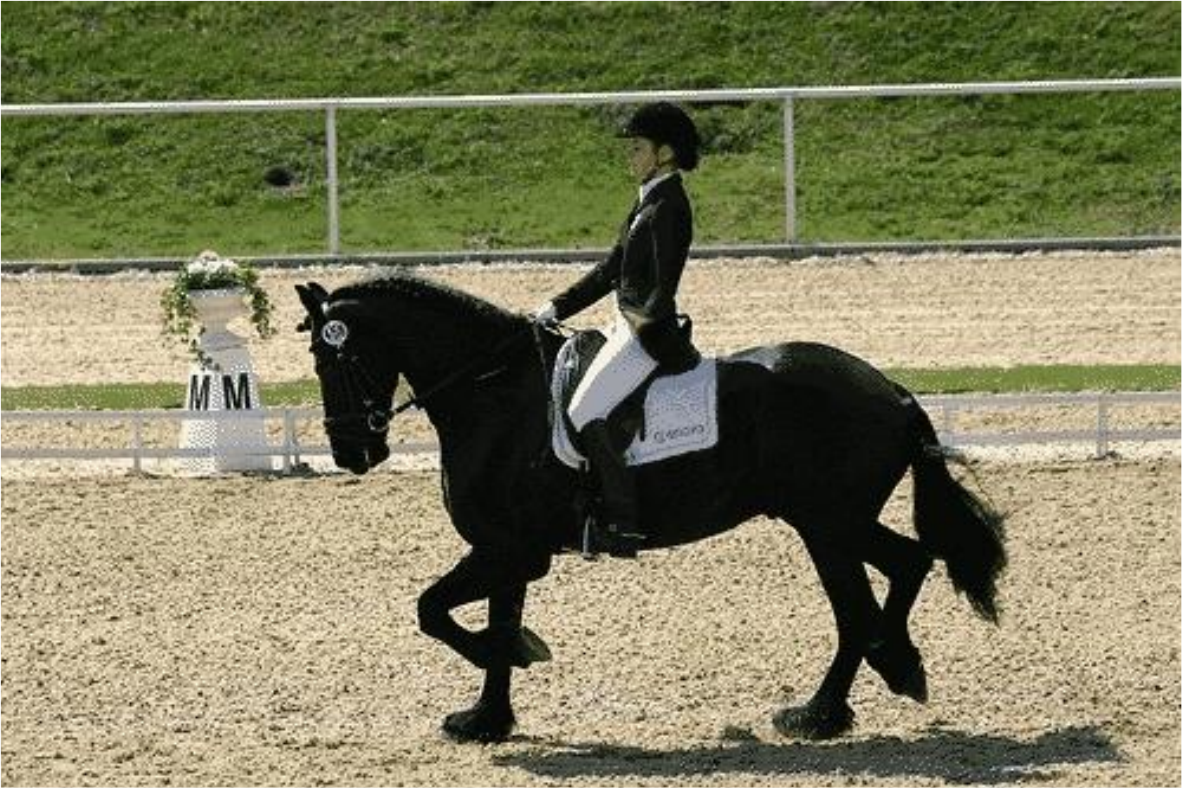}}\hskip.2em
    \subfloat[]{\includegraphics[width=0.19\textwidth]{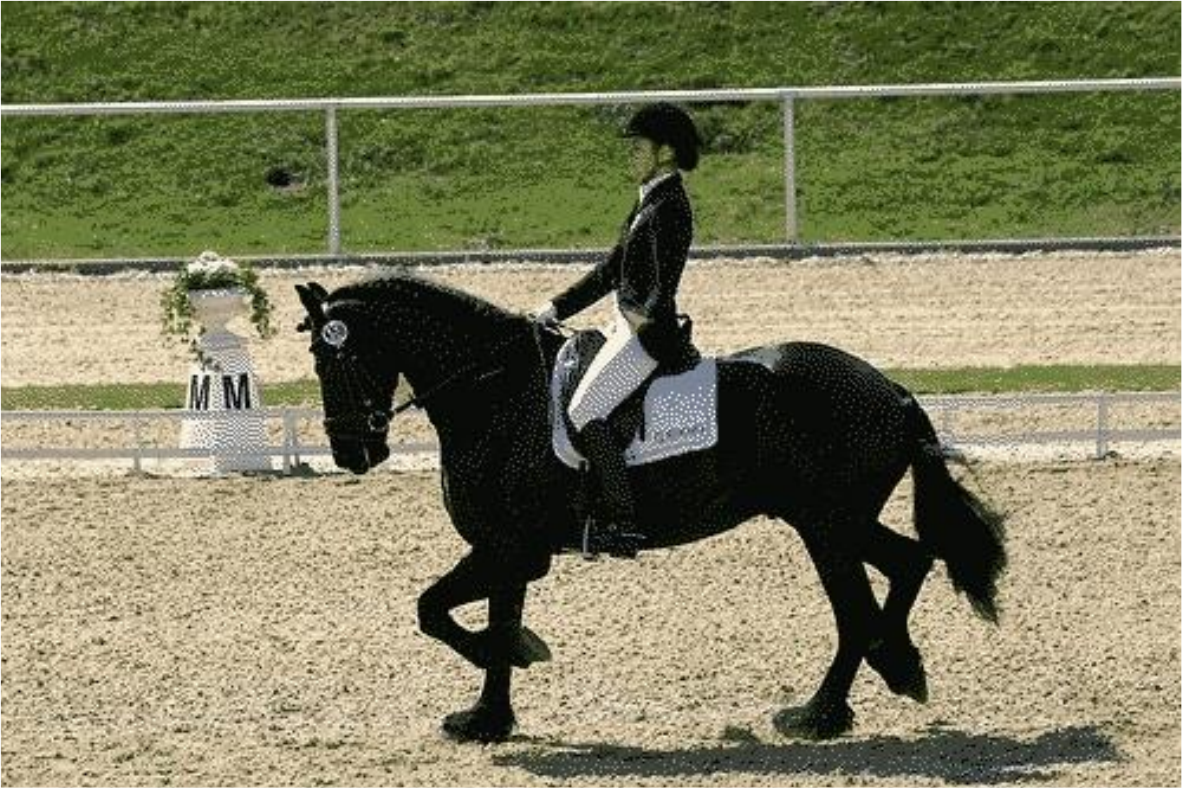}}\hskip.2em
    \subfloat[]{\includegraphics[width=0.19\textwidth]{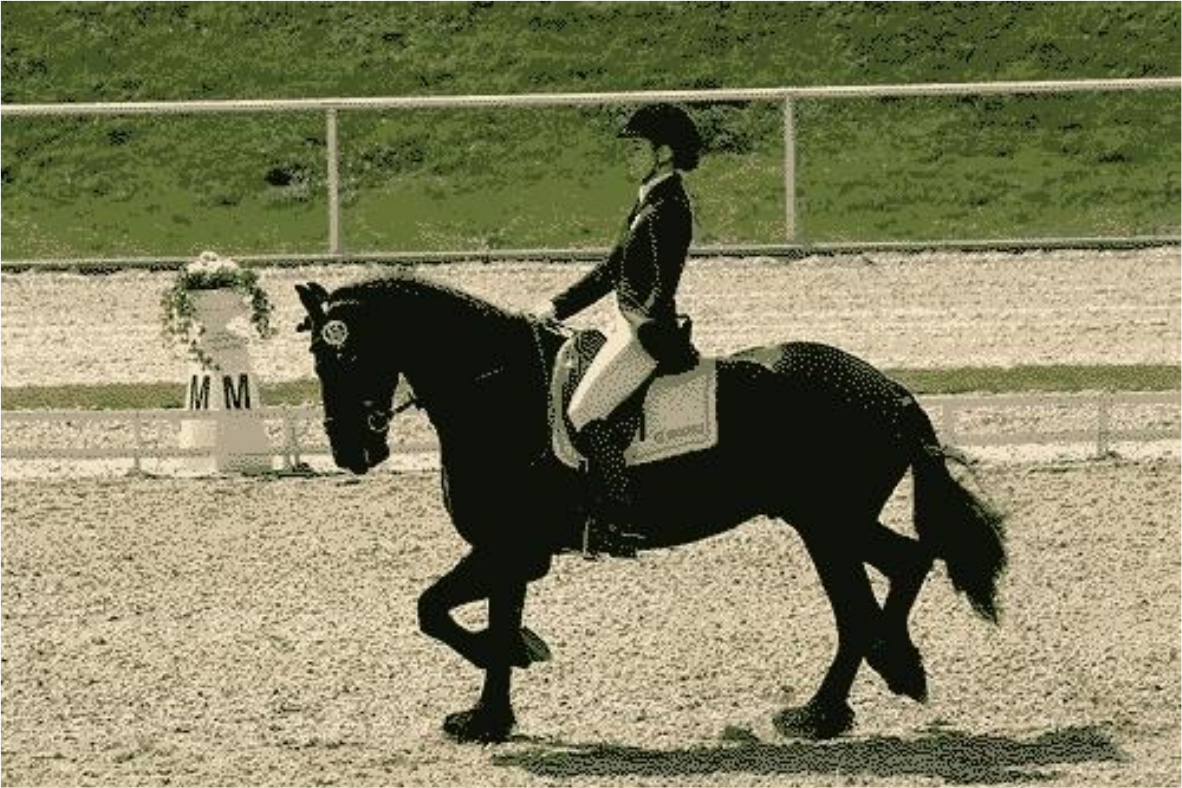}}\hskip.2em
    \subfloat[]{\includegraphics[width=0.19\textwidth]{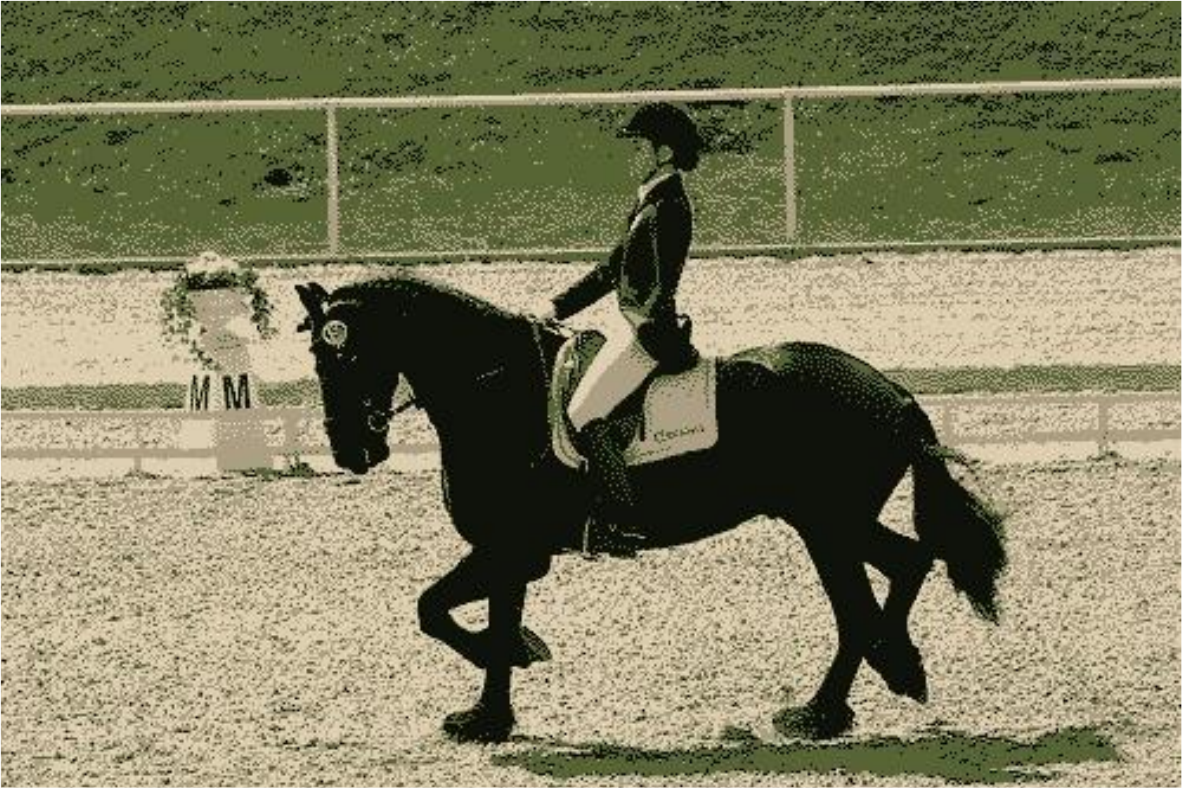}}
    \vspace{2pt}
    \subfloat[]{\includegraphics[width=0.19\textwidth]{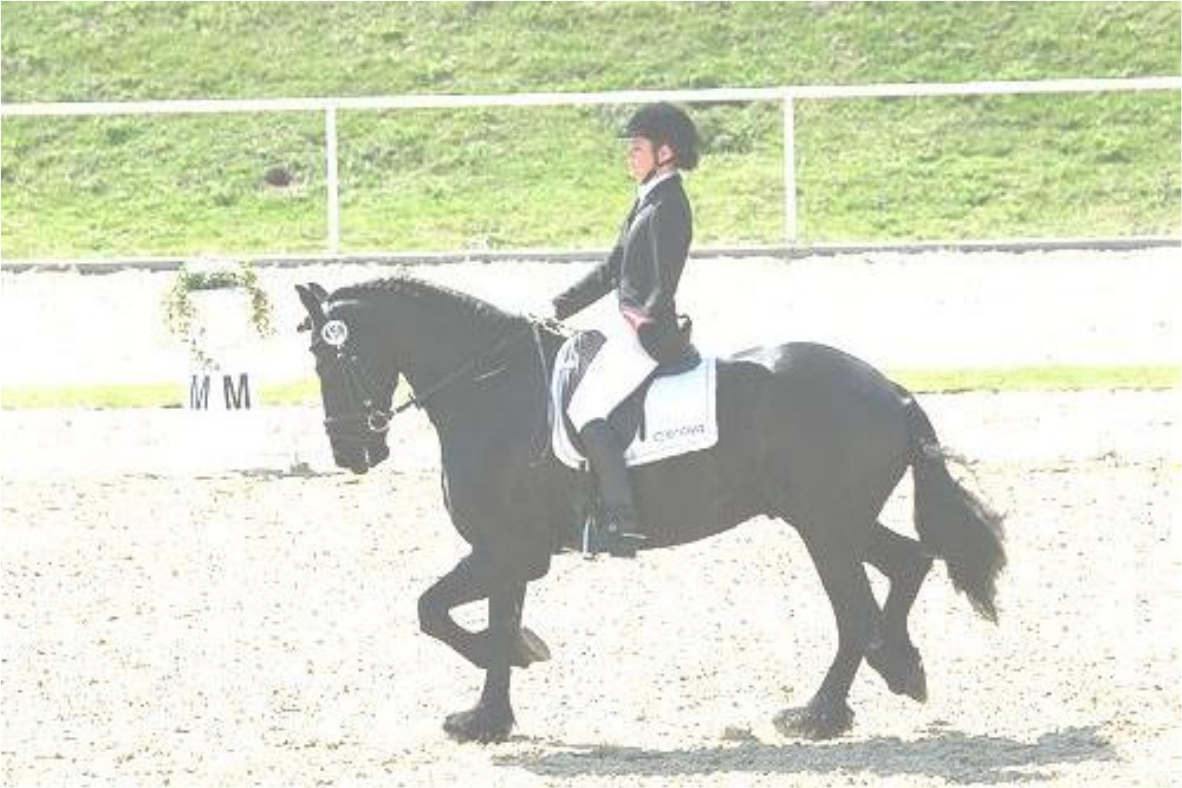}}\hskip.2em
    \subfloat[]{\includegraphics[width=0.19\textwidth]{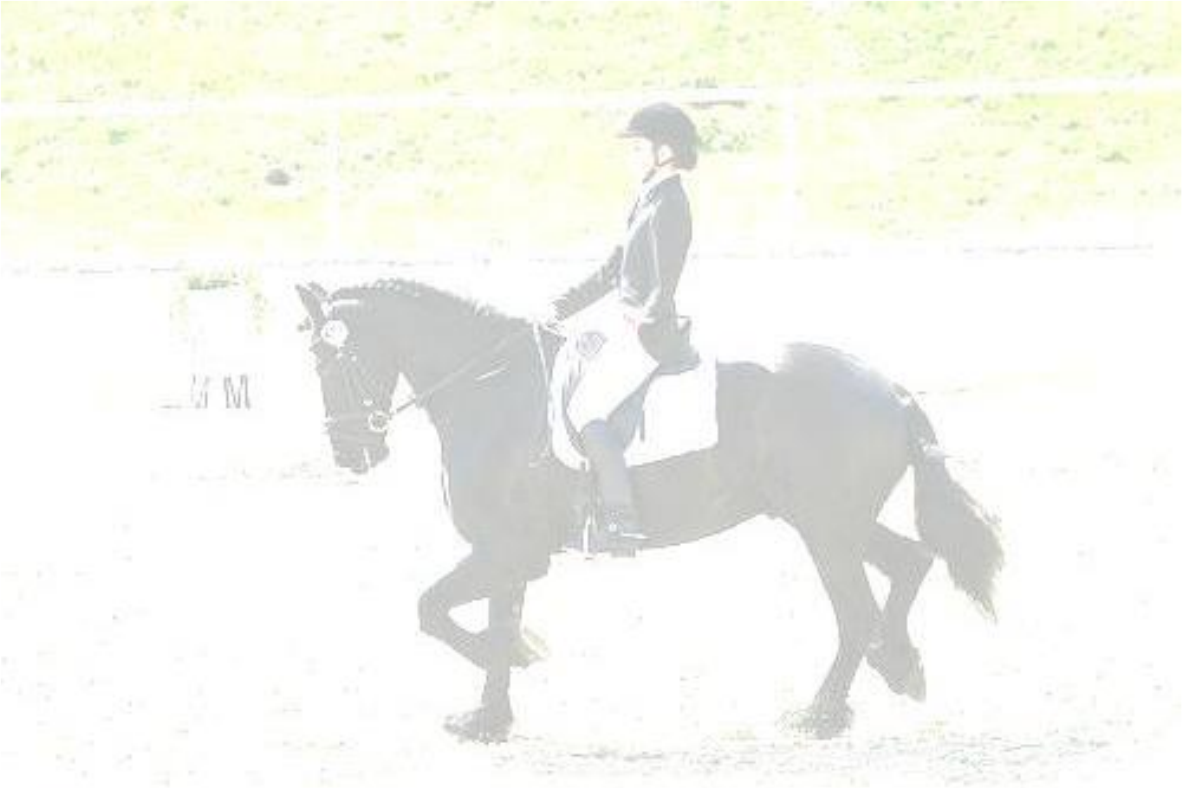}}\hskip.2em
    \subfloat[]{\includegraphics[width=0.19\textwidth]{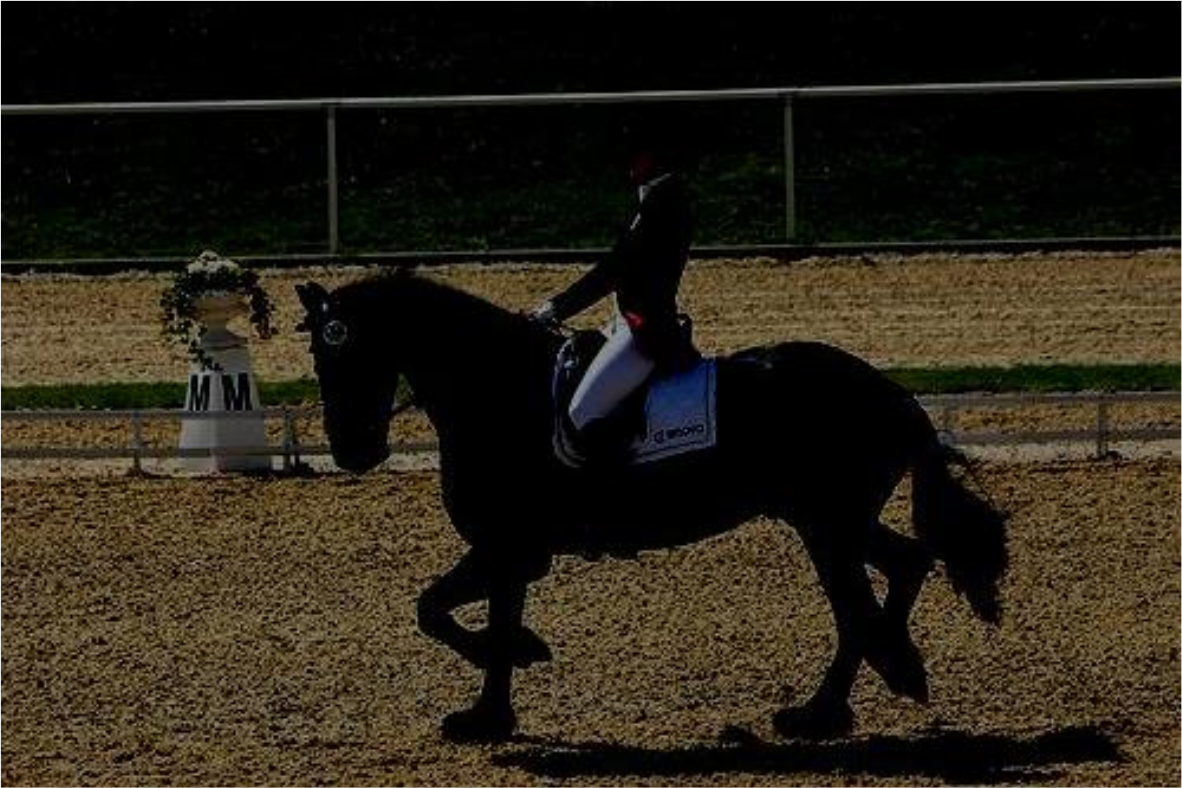}}\hskip.2em
    \subfloat[]{\includegraphics[width=0.19\textwidth]{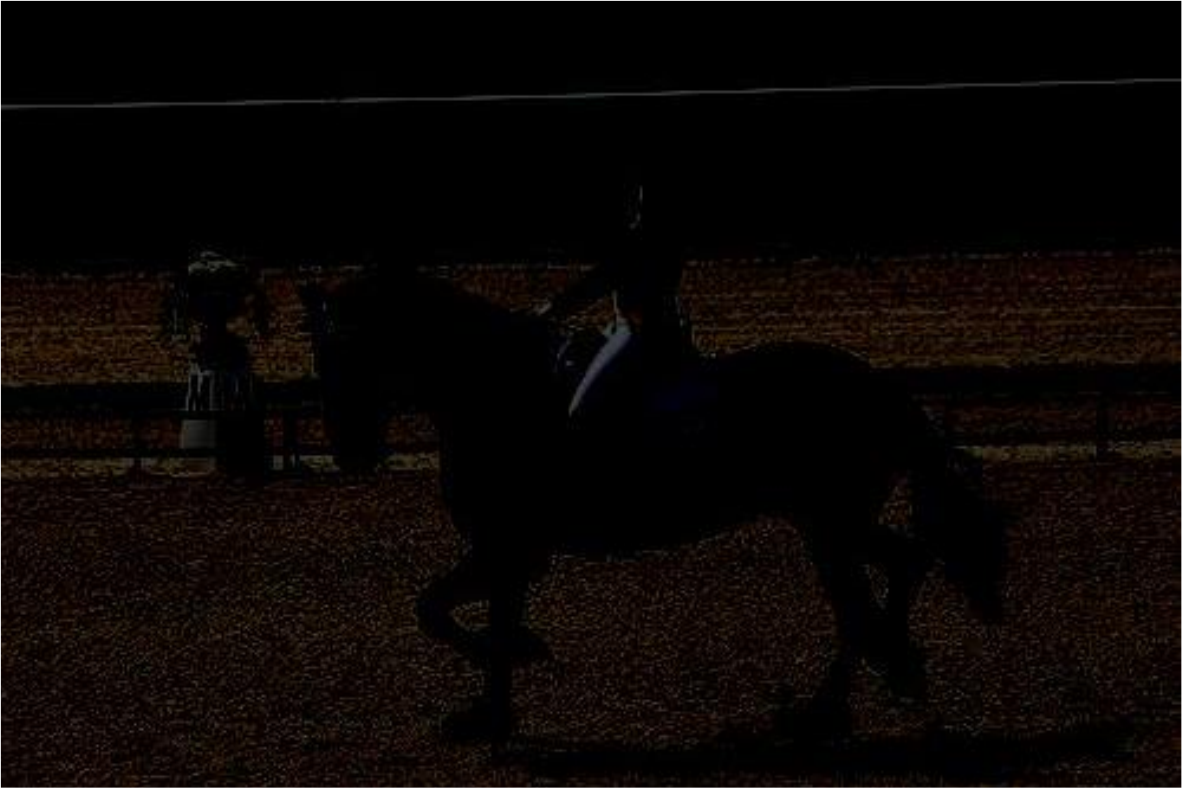}}

    \caption{Illustration of the five new distortion types with increasing  degradation levels from left to right. (a)-(e) Contrast stretching. (f)-(j) Pink noise. (k)-(o) Image color quantization with dithering. (p)-(q) Over-exposure. (r)-(s) Under-exposure.}\label{fig:degradation_level}
\end{figure*}

\section{Related Work}\label{sec:rw}
In this section, we provide a review of recent CNN-based BIQA models. For a more detailed treatment of BIQA, we refer the interested readers to~\cite{wang2011reduced,ma2017dipiq}.

Tang \textit{et al.}~\cite{tang2014blind} pre-trained a deep belief network with a radial basis function and fine-tuned it to predict image quality. Bianco \textit{et al.}~\cite{Bianco2016on} investigated
various design choices for CNN-based BIQA. They first adopted off-the-shelf
CNN features to learn a quality evaluator using support vector
regression (SVR). Alternatively, they fine-tuned the features in a multi-class classification setting followed by SVR. Their proposals are not end-to-end optimized
and involve heavy manual parameter adjustments~\cite{Bianco2016on}. Kang \textit{et al.}~\cite{kang2014convolutional} trained a CNN using a large number of spatially normalized image patches.
 Later, they estimated image quality and distortion type simultaneously via a multi-task CNN~\cite{Kang2015Simultaneous}. Patch-based training may be problematic because due to the high non-stationarity of local image content and the intricate interactions between content and distortion~\cite{Ma2018End,kim2017deep}, local image quality is not always consistent with global image quality.  Taking this problem into consideration, Bosse \textit{et al.}~\cite{bosse2016deep} trained CNN models using two strategies: direct average of features from multiple patches and weighted average of patch quality scores according to their relative importance. Kim \textit{et al.}~\cite{kim2017fully} pre-trained a CNN model using numerous patches with proxy quality scores provided by a full-reference IQA model~\cite{zhang2011fsim} and summarized the patch-level features using the mean and standard deviation statistics for fine-tuning.
A closely related work to ours is MEON~\cite{Ma2018End}, a cascaded multi-task framework for BIQA. A distortion type identification network is first trained, for which large-scale training samples are readily available. Starting from the pre-trained early layers and the outputs of the distortion type identification network, a quality prediction network is trained subsequently. Compared with MEON, the proposed DB-CNN takes a step further by considering not only distortion type but also distortion level information, which results in better quality-aware initializations. In summary, the aforementioned methods partially address the training data shortage problem in the synthetic distortion scenario, but it is difficult to extend them to the authentic distortion scenario.

\section{DB-CNN for BIQA}\label{sec:dbcnn}
In this section, we first describe the construction of the pre-training set and the CNN architecture for synthetically distorted images. We then present the tailored VGG-16 network for authentically distorted images. Finally, we introduce the bilinear pooling module along with the fine-tuning procedure.

\subsection{CNN for Synthetic Distortions}\label{subsec:scnn}
To take into account the enormous content variations in real-world images, we start with the Waterloo Exploration Database~\cite{ma2017waterloo} and the PASCAL VOC Database~\cite{everingham2010pascal}. The former contains $4,744$ pristine-quality images with four synthetic distortions, \textit{i.e.}, JPEG compression, JPEG2000 compression, Gaussian blur, and while Gaussian noise. The latter is a large database for object recognition, which contains $17,125$ images of acceptable quality with $20$ semantic classes. We merge the two databases to obtain $21,869$ source images. In addition to the four distortion types mentioned above, we add five more---contrast stretching, pink noise, image quantization with color dithering, over-exposure, and under-exposure. We ensure that the added distortions dominate the perceived quality as some source images (especially in the PASCAL VOC Database) may not have perfect quality. Following~\cite{ma2017waterloo}, we synthesize images with five distortion levels except for over-exposure and under-exposure, where only two levels are generated~\cite{Ma2015Perceptual}. Sample distorted images with various degradation levels are shown in Fig.~\ref{fig:distortions} and Fig.~\ref{fig:degradation_level}. As a result, the pre-training ？ set contains $852,891$ distorted images in total.



Due to the large scale of the pre-training set, it is impractical to carry out a full subjective experiment to obtain the MOS of each image. We take advantage of the distortion type and level information in the synthesis process, and pre-train a CNN to classify the distortion type and the degradation level. Compared to previous methods that  exploit distortion type information only~\cite{Ma2018End,Kang2015Simultaneous} , our pre-training strategy offers perceptually more meaningful initializations, leading to better local optimum (shown in Section~\ref{subsubsec:ablation}). Specifically, we form the ground truth as an $M$-class indicator vector with one entry activated to encode the underlying distortion type at specific distortion level. In our case, $M = 39$,  which corresponds to seven distortion types with five levels and two distortion types with two levels.

\begin{figure*}[t]
  \centering
  \includegraphics[width=.8\textwidth]{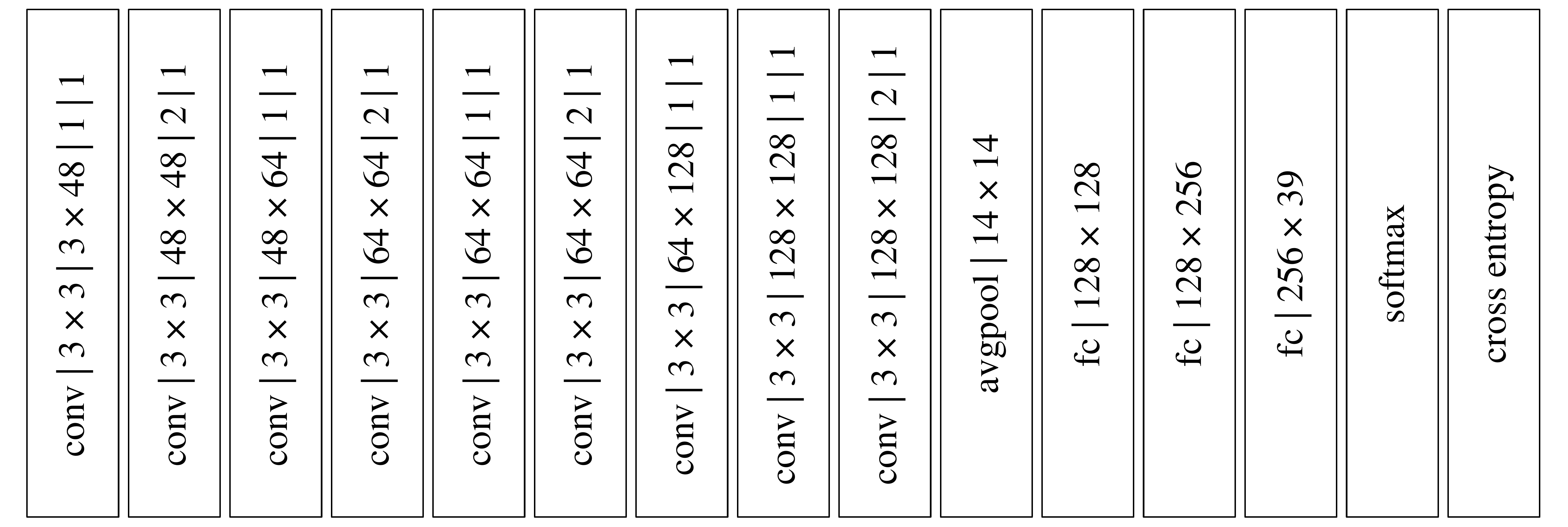}
  \caption{The architecture of S-CNN  for synthetic distortions. We follow the style and convention in~\cite{BalleLS16a}, and denote the parameterization of the convolution layer as ``height $\times$ width $\mid$ input channel $\times$ output channel $\mid$ stride $\mid$ padding''. For brevity, we ignore all ReLU layers here.}\label{fig:scnn}
\end{figure*}

Inspired by the VGG-16 network architecture~\cite{simonyan2014very}, we design our CNN for synthetic distortions (S-CNN) with a similar structure subject to some modifications (see Fig.~\ref{fig:scnn}). In a nutshell, the input image is resized and cropped to $224 \times 224\times 3$. All convolutions have a kernel size of $3\times3$ with a stride of two to reduce the spatial resolution by half in both directions.
We pad the feature activations with zeros when necessary before convolution. The nonlinear activation function we adopt is the rectified linear unit (ReLU).
The feature activations at the last convolution layer are globally averaged across spatial locations. We append three fully connected layers and the softmax layer at the end. Given $N$ training tuples ${\{(\mathbf{X}^{(1)},\mathbf{p}^{(1)}),...,(\mathbf{X}^{(N)},\mathbf{p}^{(N)})\}}$ in a mini-batch, where $\mathbf{X}^{(i)}$ denotes the $i$-th input image and $\mathbf{p}^{(i)}$ is the ground-truth indicator vector, S-CNN produces the activations of the last fully connected layer $\mathbf{y}^{(i)} = [y^{(i)}_1,\cdots, y^{(i)}_{M}]^T$. Denoting the model parameters in S-CNN by $\mathbf{W}_s$, we define the softmax function as


\begin{equation}\label{eq:softmax}
\hat{p}^{(i)}_k(\mathbf{X}^{(i)};\mathbf{W}_s) = \frac{\exp\left({y^{(i)}_k(\mathbf{X}^{(i)};\mathbf{W}_s)}\right)}{\sum_{j=1}^{M}\exp\left({y^{(i)}_j(\mathbf{X}^{(i)};\mathbf{W}_s})\right)}\,,
\end{equation}
where $\hat{\mathbf{p}}^{(i)} = [\hat{p}^{(i)}_1,\cdots, \hat{p}^{(i)}_{M}]^T$ is an $M$-dimensional probability vector of the $i$-th input, indicating the probability of each distortion type at specific degradation level. Finally, we compute the empirical cross-entropy loss by
  \begin{equation}\label{eq:cross_entropy}
\ell_s(\{\mathbf{X}^{(i)}\};\mathbf{W}_s) = -\sum_{i = 1}^{N}\sum_{j = 1}^{M}p^{(i)}_j\log \hat{p}^{(i)}_j(\mathbf{X}^{(i)};\mathbf{W}_s)\,.
\end{equation}

\subsection{CNN for Authentic Distortions}\label{subsec:bcnn}
 Unlike training S-CNN for synthetic distortions, it is difficult to obtain a large amount of relevant training data for authentic distortions. Meanwhile, training a CNN from scratch using a small number of samples often leads to overfitting. Here we resort to VGG-16~\cite{simonyan2014very} that has been pre-trained for the image classification task on ImageNet~\cite{deng2009imagenet}, to extract relevant features for authentically distorted images. Since the distortions in ImageNet occur as a natural consequence of photography rather than simulations, the VGG-16 feature representations are highly likely to adapt to authentic distortions and 	to improve the classification performance~\cite{kim2017deep}.

\subsection{DB-CNN by Bilinear Pooling}
We consider bilinear pooling to combine S-CNN for synthetic distortions and VGG-16 for authentic distortions into a unified model. Bilinear models have been shown to be effective in modeling two-factor variations, such as style and content of images~\cite{tenenbaum1997separating}, location and appearance for fine-grained recognition~\cite{lin2015bilinear}, spatial and temporal characteristics for video analysis~\cite{simonyan2014two}, and text and visual information for question-answering~\cite{fukui2016multimodal}. We tackle the BIQA problem with a similar philosophy, where synthetic and authentic distortions are modeled as two-factor variations, resulting in a DB-CNN model.

\begin{figure*}
  \centering
  \includegraphics[width=.75\textwidth]{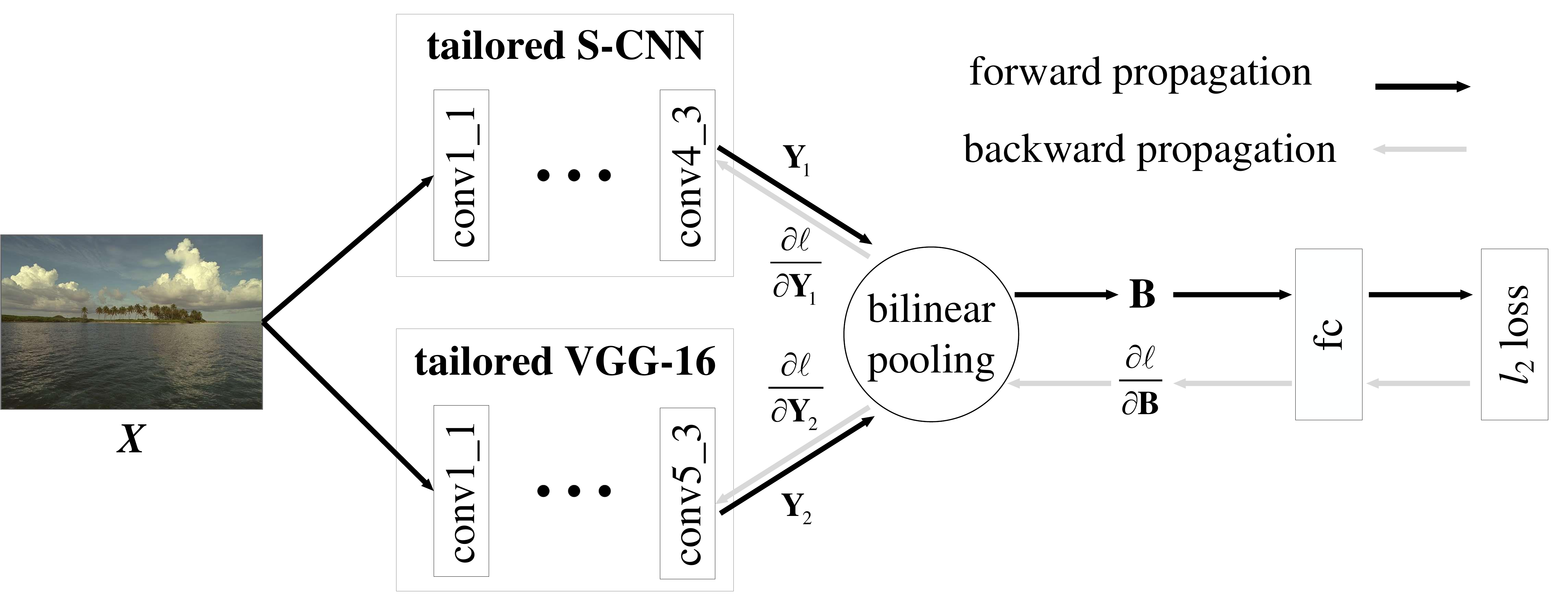}
  \caption{The structure of the proposed DB-CNN.}\label{fig:bcnn}
\end{figure*}

The structure of DB-CNN is presented in Fig.~\ref{fig:bcnn}. We tailor the pre-trained S-CNN and VGG-16 by discarding all layers after the last convolution. Denote the representations from S-CNN and VGG-16 by
$\mathbf{Y}_{1}$ and $\mathbf{Y}_{2}$, which have sizes of $h_{1} \times w_{1} \times d_{1}$ and $h_{2} \times w_{2} \times d_{2}$, respectively. The bilinear pooling of $\mathbf{Y}_{1}$ and $\mathbf{Y}_{2}$ requires $h_{1} \times w_{1} = h_{2} \times w_{2}$, which holds in our case for an input image of arbitrary size because S-CNN and VGG-16 share the same padding and downsampling routines.
Other CNNs such as ResNet~\cite{he2016deep} may also be adopted in our framework if the structure of S-CNN is adjusted appropriately. The bilinear pooling of $\mathbf{Y}_{1}$ and $\mathbf{Y}_{2}$ is formulated as
\begin{equation}\label{eq:bp}
\mathbf{B} = \mathbf{Y}_{1}^{T}\mathbf{Y}_{2},
\end{equation}
where $\mathbf{B}$ is of dimension $d_1\times d_2$. Bilinear representations are usually mapped from a Riemannian manifold into an Euclidean space~\cite{pennec2006riemannian} by
\begin{equation}\label{eq:improved}
\tilde{\mathbf{B}} = \frac{\mathrm{sign}(\mathbf{B}) \odot \sqrt{\left|\mathbf{B}\right|}}{\|\mathrm{sign}(\mathbf{B}) \odot \sqrt{\left|\mathbf{B}\right|}\|_2},
\end{equation}
where $\odot$ refers to element-wise multiplication. $\tilde{\mathbf{B}}$ is fed to a fully connected layer with one output for final quality prediction. We consider the $\ell_2$-norm as the empirical loss, which is widely used in previous studies~\cite{kang2014convolutional,bosse2016deep,kim2017deep} to drive the learning of the entire DB-CNN model on a target IQA database
\begin{equation}\label{eq:l2loss}
\ell = \frac{1}{N}\sum_{i = 1}^{N}\|s_{i} - \hat{s}_{i}\|_2,
\end{equation}
where $s_{i}$ is the MOS of the $i$-th image in a mini-batch and $\hat{s}_{i}$ is the predicted quality score by DB-CNN.

According to the chain rule, the backward propagation of the loss $\ell$ through the bilinear pooling layer to $\mathbf{Y}_{1}$ and $\mathbf{Y}_{2}$ can be computed by
\begin{equation}\label{eq:bx1}
\frac{\partial \ell}{\partial \mathbf{Y}_{1}} =  \mathbf{Y}_{2}\left(\frac{\partial \ell}{\partial \mathbf{B}}\right)^{T}
\end{equation}
and
\begin{equation}\label{eq:bx2}
\frac{\partial \ell}{\partial \mathbf{Y}_{2}} =  \mathbf{Y}_{1}\left(\frac{\partial \ell}{\partial \mathbf{B}}\right).
\end{equation}
Bilinear pooling summarizes the spatial information and enables DB-CNN to accept an input image of arbitrary size. As a result, we can feed the whole image directly instead of patches cropped from it to DB-CNN during both training and testing.

\section{Experiments}\label{sec:exp}
In this section, we first describe the experimental setups, including the IQA databases, the evaluation protocols, the performance criteria, and the implementation details of DB-CNN. After that, we compare the performance of DB-CNN with state-of-the-art BIQA models on individual databases and across databases. We also test the robustness of DB-CNN on the Waterloo Exploration Database using the discriminability and ranking consistency criteria~\cite{ma2017waterloo}, and the gMAD competition method. Finally, we conduct a series of ablation experiments to justify the rationality of DB-CNN.

\subsection{Experimental Setups}\label{subsec:expsetup}
\subsubsection{IQA Databases}\label{subsec:IQAdatabase}
The main experiments are conducted on three singly distorted synthetic IQA databases, \textit{i.e.}, LIVE~\cite{sheikh2006statistical}, CSIQ~\cite{larson2010most} and TID2013~\cite{ponomarenko2013color}, a multiply distorted synthetic dataset LIVE MD~\cite{Jayaraman2013Objective}, and the authentic LIVE Challenge Database~\cite{ghadiyaram2016massive}. LIVE~\cite{sheikh2006statistical} contains $779$ distorted images synthesized from $29$ reference images with five distortion types---JPEG compression (JPEG), JPEG2000 compression (JP2K), Gaussian blur (GB), white Gaussian noise (WN), and fast fading error (FF) at seven to eight degradation levels. Difference MOS (DMOS) in the range of $[0, 100]$ is collected for each image with a higher value indicating lower perceptual quality. CSIQ~\cite{larson2010most}  is composed of $866$ distorted images generated from $30$ reference images, including six distortion types, \textit{i.e.}, JPEG, JP2K, GB, WN, contrast change (CG), and pink noise (PN) at three to five degradation levels. DMOS in the range of $[0, 1]$ is provided as the ground truth. TID2013~\cite{ponomarenko2013color} consists of $3,000$ distorted images from $25$ reference images with $24$ distortion types at five degradation levels. MOS in the range of $[0, 9]$ is provided to indicate perceptual quality. LIVE MD~\cite{Jayaraman2013Objective} contains 450 images generated from 15 source images under two multiple distortion scenarios---blur followed by JPEG compression and blur followed by white Gaussian noise. DMOS in the range of $[0, 100]$ is provided as the subjective opinion. LIVE Challenge~\cite{ghadiyaram2016massive} is an authentic IQA database, which contains $1,162$ images captured from diverse real-world scenes by numerous photographers with various levels of photography skills using different camera devices. As a result, the images undergo complex realistic distortions. MOS in the range of $[0, 100]$ is collected from over $8,100$ unique human evaluators via an online crowdsourcing platform.

\subsubsection{Experimental Protocols and Performance Criteria}\label{subsec:expprotocol}
We conduct the experiments by following the same protocol in~\cite{kim2017deep}. Specifically, we divide the distorted images in a target IQA database into two splits, $80\%$ of which are used for fine-tuning  DB-CNN and the rest $20\%$ for testing. For synthetic databases LIVE, CSIQ, TID2013, and LIVE MD, we guarantee the image content independence between the fine-tuning and test sets. The splitting procedure is randomly repeated ten times for all databases and the average results are reported.

We adopt two commonly used metrics to benchmark BIQA models: Spearman rank correlation coefficient (SRCC) and Pearson linear correlation coefficient (PLCC). SRCC measures the prediction monotonicity while PLCC measures the prediction precision. As suggested in~\cite{video2003final}, the predicted quality scores are passed through a nonlinear logistic function before computing PLCC
\begin{equation}\label{eq:logistic}
\tilde{s} = \beta_{1}\left(\frac{1}{2}-\frac{1}{\exp(\beta_{2}(\hat{s}-\beta_{3}))}\right)+\beta_{4}\hat{s}+\beta_{5},
\end{equation}
where $\{\beta_{i}; i=1,2,3,4,5\}$ are regression parameters to be fitted.

\subsubsection{Implementation Details}\label{subsec:impldetail}
All parameters in S-CNN are initialized by He's method~\cite{he2015delving} and trained from scratch using Adam~\cite{Kingma2014adam} with a mini-batch of $64$.  We run $30$ epochs with a learning rate decaying logarithmically from $10^{-3}$ to $10^{-5}$. Images are first scaled to $256 \times 256 \times 3$ and cropped to $224 \times 224 \times 3$ as inputs. During fine-tuning of DB-CNN, we adopt Adam~\cite{Kingma2014adam} again with a learning rate of $10^{-6}$ for LIVE~\cite{sheikh2006statistical} and CSIQ~\cite{larson2010most}, and $10^{-5}$ for TID2013~\cite{ponomarenko2013color}, LIVE MD~\cite{Jayaraman2013Objective} and LIVE Challenge~\cite{ghadiyaram2016massive}, respectively. The mini-batch size is set to eight. Batch normalization~\cite{ioffe2015batch} is used to stabilize the pre-training and fine-tuning.
We feed images of original size to DB-CNN during both fine-tuning and test phases. We implement DB-CNN using the MatConvNet toolbox~\cite{vedaldi2015matconvnet} and will release the code at \url{https://github.com/zwx8981/BIQA_project}.

\begin{table}[t]
  \centering
  \caption{Average SRCC and PLCC results across ten sessions. The top two results are highlighted in boldface. LIVE CL stands for the LIVE Challenge Database}\label{tab:overall}
  \begin{tabular}{l|ccccc}
      \toprule
        \multirow{2}{*}{SRCC} & LIVE & CSIQ& TID2013 & LIVE & LIVE\\
         & \cite{sheikh2006statistical} & \cite{larson2010most} & \cite{ponomarenko2013color} & MD \cite{Jayaraman2013Objective} & CL\cite{ghadiyaram2016massive}\\
     \hline

      BRISQUE~\cite{mittal2012no} & 0.939 & 0.746 & 0.604 & 0.886 & 0.608 \\
      M3~\cite{xue2014blind}& 0.951 & 0.795 & 0.689 & 0.892 & 0.607 \\
      FRIQUEE~\cite{ghadiyaram2017perceptual} & 0.940 & 0.835 & 0.680 & {\bf 0.923} & 0.682\\
      CORNIA~\cite{ye2012unsupervised} & 0.947 & 0.678 & 0.678 & 0.899 & 0.629 \\
      HOSA~\cite{xu2016blind} & 0.946 & 0.741 & 0.735 & 0.913 & 0.640 \\
      Le-CNN~\cite{kang2014convolutional} &0.956 & --- & ---& --- & --- \\
            BIECON~\cite{kim2017fully} & 0.961 & 0.815 & 0.717 & 0.909 & 0.595 \\
            DIQaM~\cite{bosse2016deep} & 0.960 & --- & {\bf 0.835} & --- & 0.606 \\
            WaDIQaM~\cite{bosse2016deep} & 0.954 & --- & 0.761 & --- & 0.671 \\
            ResNet-ft~\cite{kim2017deep} & 0.950 & {\bf 0.876} & 0.712 & 0.909 &{\bf 0.819} \\
            IW-CNN~\cite{kim2017deep} & {\bf 0.963} & 0.812 & 0.800 & 0.914 & 0.663 \\
     \hline
        DB-CNN& {\bf 0.968} & {\bf 0.946}& {\bf 0.816} & {\bf 0.927} &{\bf 0.851} \\
   \midrule

                \multirow{2}{*}{PLCC} & \multirow{2}{*}{LIVE} & \multirow{2}{*}{CSIQ}& \multirow{2}{*}{TID2013} & LIVE & LIVE\\
         &  &  & & MD  & CL\\
     \hline

      BRISQUE~\cite{mittal2012no} & 0.935 & 0.829 & 0.694 & 0.917 & 0.629  \\
      M3~\cite{xue2014blind}& 0.950 & 0.839 & 0.771 & 0.919 & 0.630 \\
      FRIQUEE~\cite{ghadiyaram2017perceptual} & 0.944 & 0.874 & 0.753 & {\bf 0.934} & 0.705\\
      CORNIA~\cite{ye2012unsupervised} & 0.950 & 0.776 & 0.768 & 0.921 & 0.671 \\
      HOSA~\cite{xu2016blind} & 0.947 & 0.823 & 0.815 & 0.926 & 0.678 \\
      Le-CNN~\cite{kang2014convolutional} &0.953 & --- & --- & --- & --- \\
            BIECON~\cite{kim2017fully} & 0.962 & 0.823 & 0.762 & 0.933 & 0.613 \\
            DIQaM~\cite{bosse2016deep} & {\bf 0.972} & --- & {\bf 0.855} & --- & 0.601 \\
            WaDIQaM~\cite{bosse2016deep} & 0.963 & --- & 0.787 & --- & 0.680 \\
            ResNet-ft~\cite{kim2017deep} & 0.954 & {\bf 0.905} & 0.756 & 0.920 & {\bf 0.849} \\
            IW-CNN~\cite{kim2017deep} & 0.964 & 0.791 & 0.802 & 0.929 & 0.705 \\
     \hline
        DB-CNN& {\bf 0.971} & {\bf 0.959} & {\bf 0.865} & {\bf 0.934} &{\bf 0.869} \\
     \bottomrule
   \end{tabular}
\end{table}

\subsection{Experimental Results}\label{subsec:exp results}
\subsubsection{ Performance on Individual Databases}\label{subsubsec:overall_results}
We compare DB-CNN against several state-of-the-art BIQA models. The source codes of BRISQUE~\cite{mittal2012no}, M3~\cite{xue2014blind}, FRIQUEE~\cite{ghadiyaram2017perceptual}, CORNIA~\cite{ye2012unsupervised}, HOSA~\cite{xu2016blind}, and dipIQ~\cite{ma2017dipiq} are provided by the respective authors. We re-train and/or validate them using the same randomly generated training-test splits.  For CNN-based counterparts, we directly copy the performance from the corresponding papers due to the unavailability of the training codes. The SRCC and PLCC results on the five databases are listed in Table~\ref{tab:overall}, from which we have several interesting observations.  First, while all competing models achieve comparable performance on LIVE~\cite{sheikh2006statistical}, their results on CSIQ~\cite{larson2010most} and TID2013~\cite{ponomarenko2013color} are rather diverse. Compared with knowledge-driven models, CNN-based models deliver better performance on CSIQ and TID2013 because of end-to-end feature learning rather than hand-crafted feature engineering.  Second, on the multiply distorted dataset LIVE MD, DB-CNN performs favorably although it does not include multiply distorted images for pre-training, indicating that DB-CNN generalizes well to slightly different distortion scenarios. Last, for the authentic database LIVE Challenge, FRIQUEE~\cite{ghadiyaram2017perceptual} that combines a set of quality-aware features extracted from multiple color spaces outperforms other knowledge-driven BIQA models and all CNN-based models except for ResNet-ft~\cite{kim2017deep} and the proposed DB-CNN. This suggests that the intrinsic characteristics of authentic distortions cannot be fully captured by low-level features learned from synthetically distorted images. The success of DB-CNN on LIVE Challenge verifies the relevance between the high-level features from VGG-16 and the authentic distortions. In summary, DB-CNN achieves superior performance on both synthetic and authentic IQA databases.

\begin{table}[t]
  \centering
  \caption{Average SRCC and PLCC results of individual distortion types across ten sessions on LIVE~\cite{sheikh2006statistical}}\label{tab:live_idd}
  \begin{tabular}{l|ccccc}
      \toprule
        SRCC & JPEG & JP2K & WN & GB & FF \\
     \hline

      BRISQUE~\cite{mittal2012no} & 0.965 & 0.929 & 0.982 & {\bf 0.964} & 0.828 \\
      M3~\cite{xue2014blind}& 0.966 & 0.930 & {\bf 0.986} & 0.935 & 0.902 \\
      FRIQUEE~\cite{ghadiyaram2017perceptual} & 0.947 & 0.919 & {\bf 0.983} & 0.937 & 0.884 \\
      CORNIA~\cite{ye2012unsupervised} & 0.947 & 0.924 & 0.958 & 0.951 & 0.921 \\
      HOSA~\cite{xu2016blind} & 0.954 & 0.935 & 0.975 & {\bf 0.954} & {\bf 0.954} \\
            dipIQ~\cite{ma2017dipiq} & {\bf 0.969} & {\bf 0.956} & 0.975 & 0.940 & --- \\
     \hline
        DB-CNN& {\bf 0.972} & {\bf 0.955} & 0.980 & 0.935 & {\bf 0.930} \\
   \midrule
        PLCC & JPEG & JP2K & WN & GB &  FF \\
     \hline

      BRISQUE~\cite{mittal2012no} & 0.971 & 0.940 & 0.989 & {\bf 0.965} & 0.894 \\
      M3~\cite{xue2014blind}& 0.977 & 0.945 & {\bf 0.992} & 0.947 & 0.920 \\
      FRIQUEE~\cite{ghadiyaram2017perceptual} & 0.955 & 0.935 & {\bf 0.991} & 0.949 & 0.936 \\
      CORNIA~\cite{ye2012unsupervised} & 0.962 & 0.944 & 0.974 & 0.961 & 0.943 \\
      HOSA~\cite{xu2016blind} & 0.967 & 0.949 & 0.983 & {\bf 0.967} & {\bf 0.967} \\
            dipIQ~\cite{ma2017dipiq} & {\bf 0.980} & {\bf 0.964} & 0.983 & 0.948 & --- \\
     \hline
        DB-CNN& {\bf 0.986} & {\bf 0.967} & 0.988 & 0.956 & {\bf 0.961} \\
     \bottomrule
   \end{tabular}
\end{table}

\begin{table}[t]
  \centering
  \caption{Average SRCC and PLCC results of individual distortion types across ten sessions on CSIQ~\cite{larson2010most}}\label{tab:csiq_idd}
  \begin{tabular}{l|cccccc}
      \toprule
        SRCC & JPEG & JP2K & WN & GB & PN & CC \\
     \hline

      BRISQUE~\cite{mittal2012no} & 0.806 & 0.840 & 0.723 & 0.820 & 0.378 & 0.804 \\
      M3~\cite{xue2014blind}& 0.740 & 0.911 & 0.741 & 0.868 & 0.663 & 0.770 \\
      FRIQUEE~\cite{ghadiyaram2017perceptual} & 0.869 & 0.846 & 0.748 & 0.870 & {\bf 0.753} & {\bf 0.838} \\
      CORNIA~\cite{ye2012unsupervised} & 0.513 & 0.831 & 0.664 & 0.836 & 0.493 & 0.462 \\
      HOSA~\cite{xu2016blind} & 0.733 & 0.818 & 0.604 & 0.841 & 0.500 & 0.716 \\
            dipIQ~\cite{ma2017dipiq} & 0.936 & {\bf 0.944} & 0.904 & {\bf 0.932} & --- & --- \\
            MEON~\cite{Ma2018End} & {\bf 0.948} & 0.898 & {\bf 0.951} & 0.918 & --- & --- \\
     \hline
        DB-CNN& {\bf 0.940} & {\bf 0.953} & {\bf 0.948} & {\bf 0.947} & {\bf 0.940} & {\bf 0.870} \\
   \midrule
        PLCC & JPEG & JP2K & WN & GB & PN & CC \\
     \hline

      BRISQUE~\cite{mittal2012no} & 0.828 & 0.887 & 0.742 & 0.891 & 0.496 & 0.835 \\
      M3~\cite{xue2014blind}& 0.768 & 0.928 & 0.728 & 0.917 & 0.717 & 0.787 \\
      FRIQUEE~\cite{ghadiyaram2017perceptual} & 0.885 & 0.883 & 0.778 & 0.905 & {\bf 0.769} & {\bf 0.864} \\
      CORNIA~\cite{ye2012unsupervised} & 0.563 & 0.883 & 0.687 & 0.904 & 0.632 & 0.543 \\
      HOSA~\cite{xu2016blind} & 0.759 & 0.899 & 0.656 & 0.912 & 0.601 & 0.744 \\
            dipIQ~\cite{ma2017dipiq} & 0.975 & {\bf 0.959} & 0.927 & {\bf 0.958} & --- & --- \\
            MEON~\cite{Ma2018End} & {\bf 0.979} & 0.925 & {\bf 0.958} & 0.946 & --- & --- \\
     \hline
        DB-CNN& {\bf 0.982} & {\bf 0.971} & {\bf 0.956} & {\bf 0.969} & {\bf 0.950} & {\bf 0.895} \\
     \bottomrule
   \end{tabular}
\end{table}

\begin{table*}[t]
  \centering
  \caption{Average SRCC results of individual distortion types across ten sessions on TID2013~\cite{ponomarenko2013color}. We obtain similar results using PLCC, which are omitted here due to the page limit}\label{tab:tid_idd}
  \begin{tabular}{l|cccccc|c}
      \toprule
        SRCC & BRISQUE~\cite{mittal2012no} & M3~\cite{xue2014blind} & FRIQUEE~\cite{ghadiyaram2017perceptual} &
            CORNIA~\cite{ye2012unsupervised} & HOSA~\cite{xu2016blind} & MEON~\cite{Ma2018End} & DB-CNN \\
     \hline

      Additive Gaussian noise & 0.711 & 0.766 & 0.730 & 0.692 & {\bf 0.833} & {\bf 0.813} & 0.790 \\
      Additive noise in color components & 0.432 & 0.560 & 0.573 & 0.137 & 0.551 & {\bf 0.722} & {\bf 0.700} \\
        Spatially correlated noise & 0.746 & 0.782 & {\bf 0.866} & 0.741 & 0.842 & {\bf 0.926} & 0.826 \\
      Masked noise & 0.252 & 0.577 & 0.345 & 0.451 & 0.468 & {\bf 0.728} & {\bf 0.646} \\
      High frequency noise & 0.842 & {\bf 0.900} & 0.847 & 0.815 & 0.897 & {\bf 0.911} & 0.879 \\
      Impulse noise & 0.765 & 0.738 & 0.730 & 0.616 & {\bf 0.809} & {\bf 0.901} & 0.708 \\
            Quantization noise & 0.662 & {\bf 0.832} & 0.764 & 0.661 & 0.815 & {\bf 0.888} & 0.825 \\
            Gaussian blur & 0.871 & {\bf 0.896} & 0.881 & 0.850 & 0.883 & {\bf 0.887} & 0.859 \\
            Image denoising & 0.612 & 0.709 & 0.839 & 0.764 & {\bf 0.854} & 0.797 & {\bf 0.865} \\
            JPEG compression & 0.764 & 0.844 & 0.813 & 0.797 & {\bf 0.891} & 0.850 & {\bf 0.894} \\
            JPEG2000 compression & 0.745 & 0.885 & 0.831 & 0.846 & {\bf 0.919} & 0.891 & {\bf 0.916} \\
            JPEG transmission errors & 0.301 & 0.375 & 0.498 & 0.694 & 0.730 & {\bf 0.746} & {\bf 0.772} \\
            JPEG2000 transmission errors & {\bf 0.748} & 0.718 & 0.660 & 0.686 & 0.710 & 0.716 & {\bf 0.773} \\
            Non-eccentricity pattern noise & {\bf 0.269} & 0.173 & 0.076 & 0.200 & 0.242 & 0.116 & {\bf 0.270} \\
            Local bock-wise distortions & 0.207 & 0.379 & 0.032 & 0.027 & 0.268 & {\bf 0.500} & {\bf 0.444} \\
            Mean shift & 0.219 & 0.119 & {\bf 0.254} & {\bf 0.232}  & 0.211 & 0.177 & -0.009 \\
            Contrast change & -0.001 & 0.155 & {\bf 0.585} & 0.254 & 0.362 & 0.252 & {\bf 0.548} \\
            Change of color saturation & 0.003 & -0.199 & 0.589 & 0.169 & 0.045 & {\bf 0.684} & {\bf 0.631} \\
            Multiplicative Gaussian noise & 0.717 & 0.738 & 0.704 & 0.593 & {\bf 0.768} & {\bf 0.849} & 0.711 \\
            Comfort noise & 0.196 & 0.353 & 0.318 & 0.617 & {\bf 0.622} & 0.406 & {\bf 0.752} \\
            Lossy compression of noisy images & 0.609 & 0.692 & 0.641 & 0.712 & {\bf 0.838} & 0.772 & {\bf 0.860} \\
            Color quantization with dither & 0.831 & {\bf 0.908} & 0.768 & 0.683 & {\bf 0.896} & 0.857 & 0.833 \\
            Chromatic aberrations & 0.615 & 0.570 & 0.737 & 0.696 & {\bf 0.753} & {\bf 0.779} & 0.732 \\
            Sparse sampling and reconstruction  & 0.807 & 0.893 & 0.891 & 0.865 & {\bf 0.909} & 0.855 & {\bf 0.902} \\
     \bottomrule
   \end{tabular}
\end{table*}

\begin{figure*}
    \centering
    \captionsetup{justification=centering}

    \subfloat[]{\includegraphics[width=0.32\textwidth]{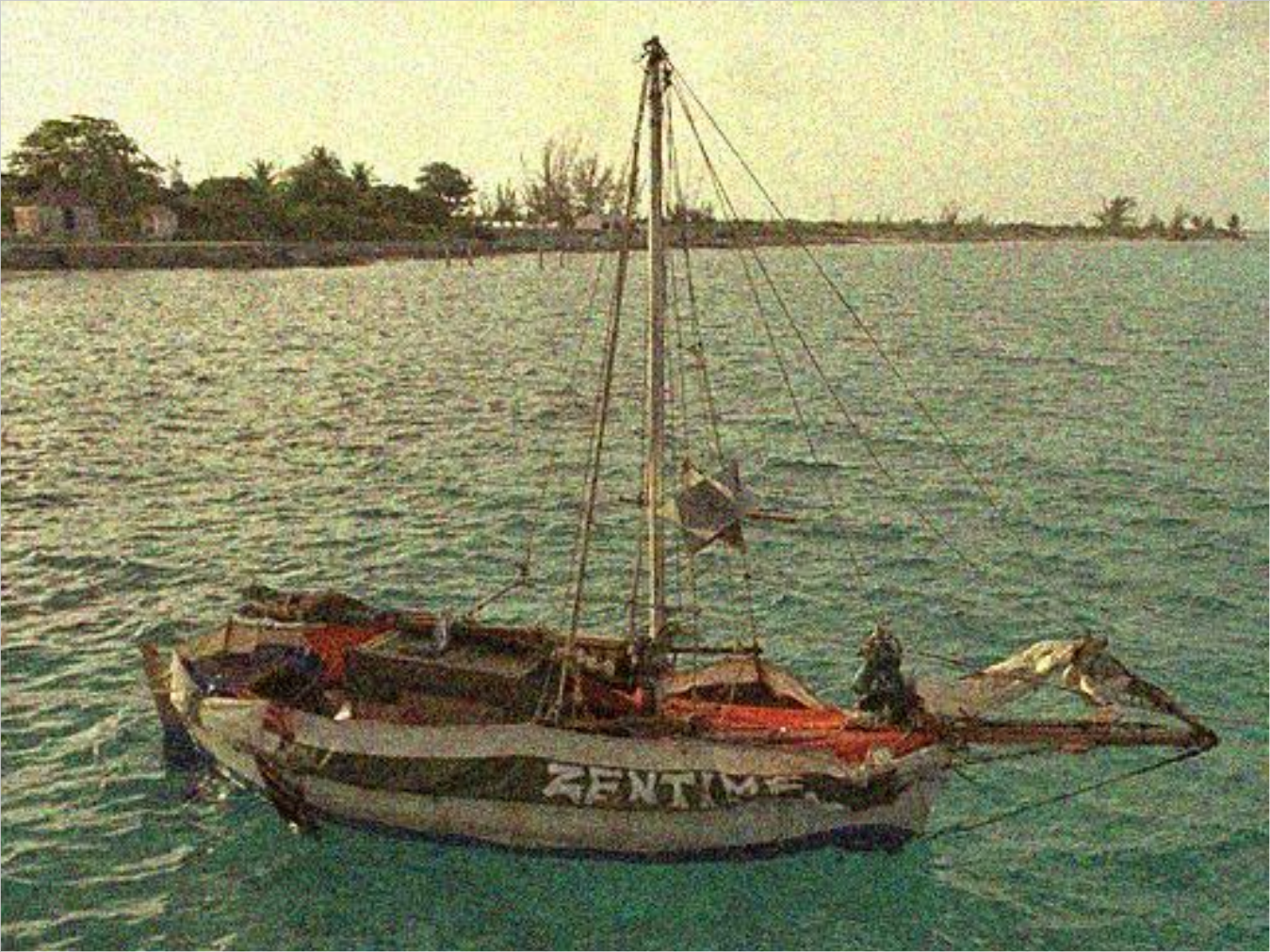}}\hskip.2em
    \subfloat[]{\includegraphics[width=0.32\textwidth]{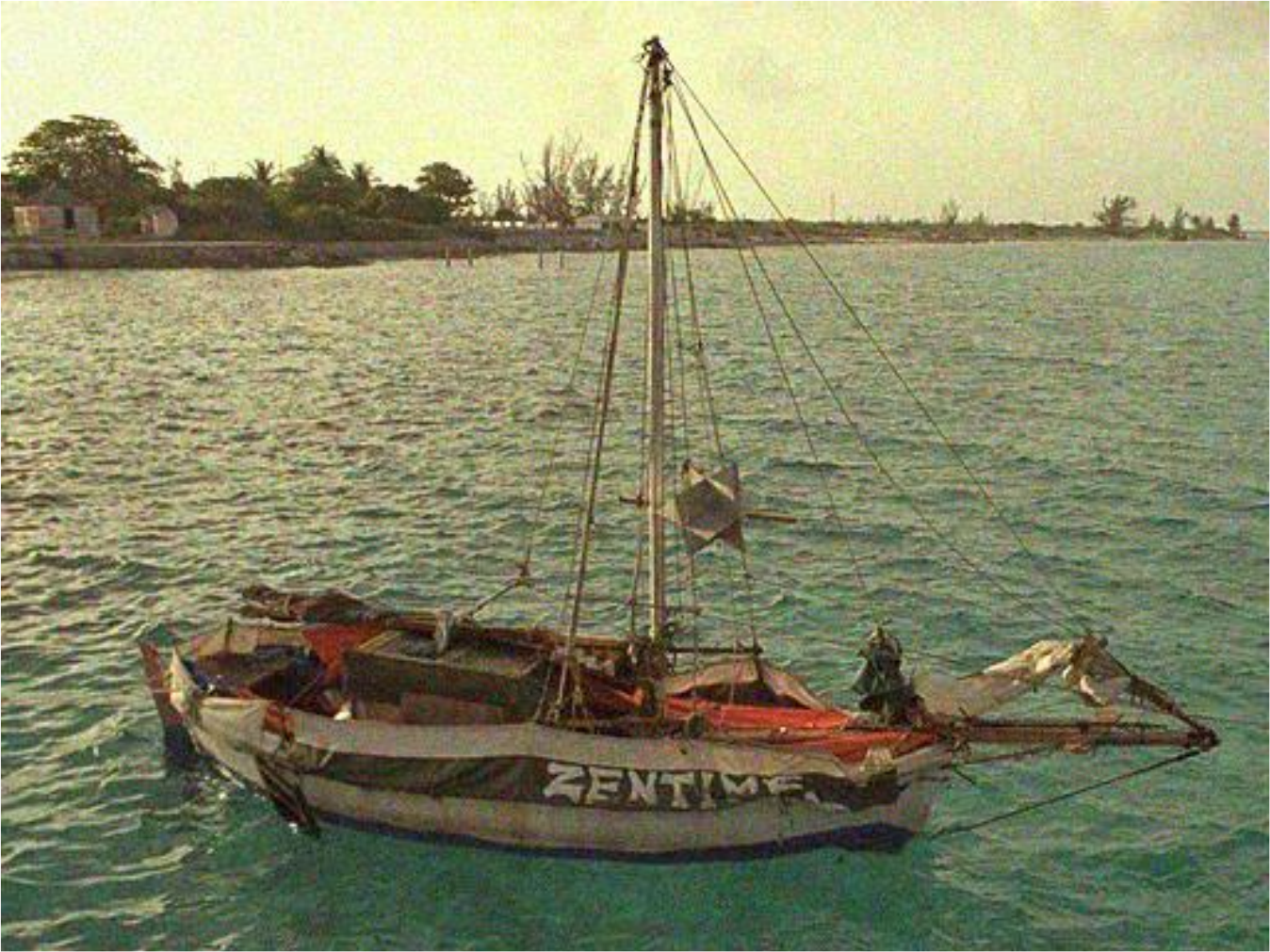}}\hskip.2em
    \subfloat[]{\includegraphics[width=0.32\textwidth]{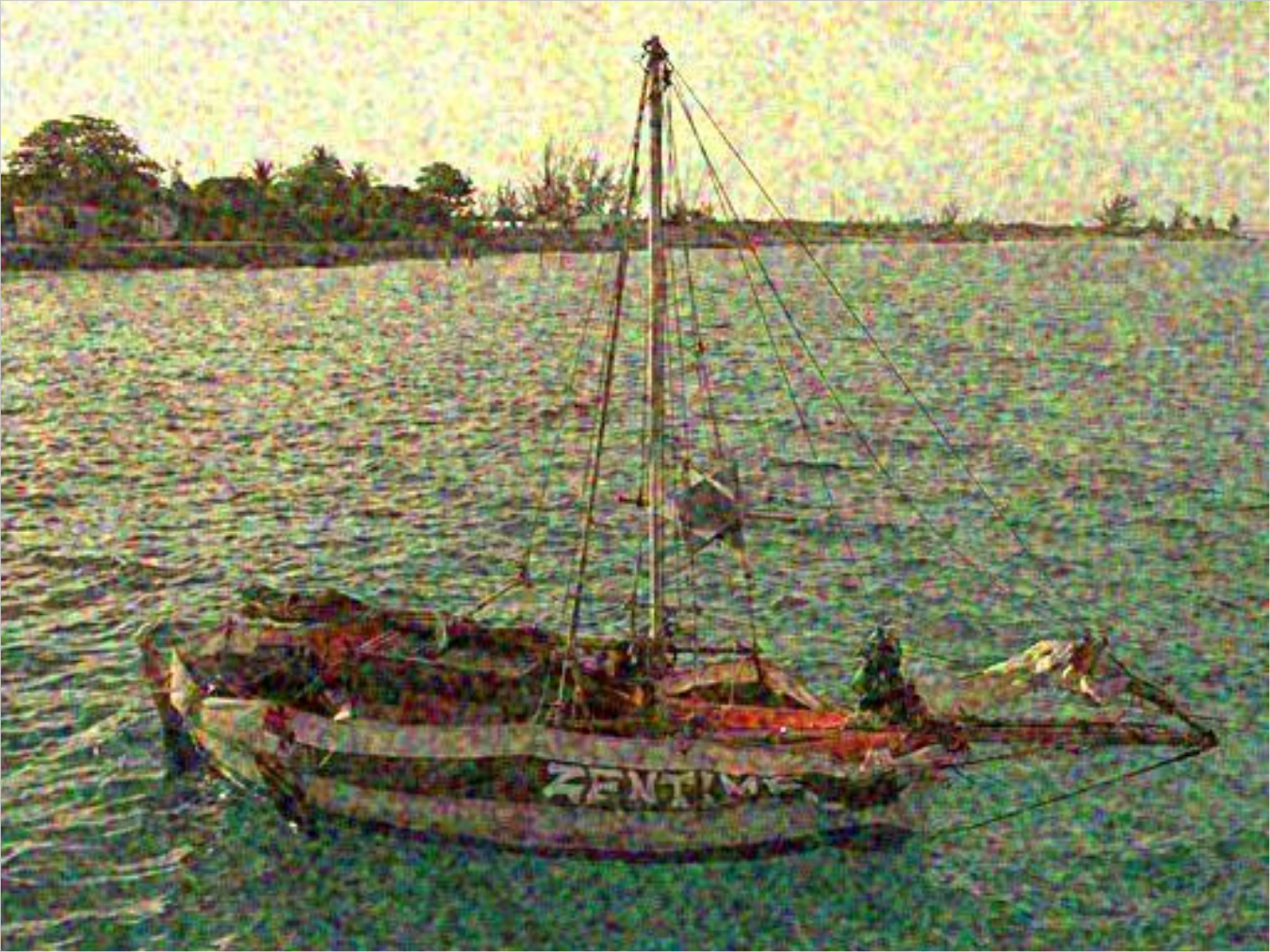}}
    \vspace{2pt}
    \subfloat[]{\includegraphics[width=0.32\textwidth]{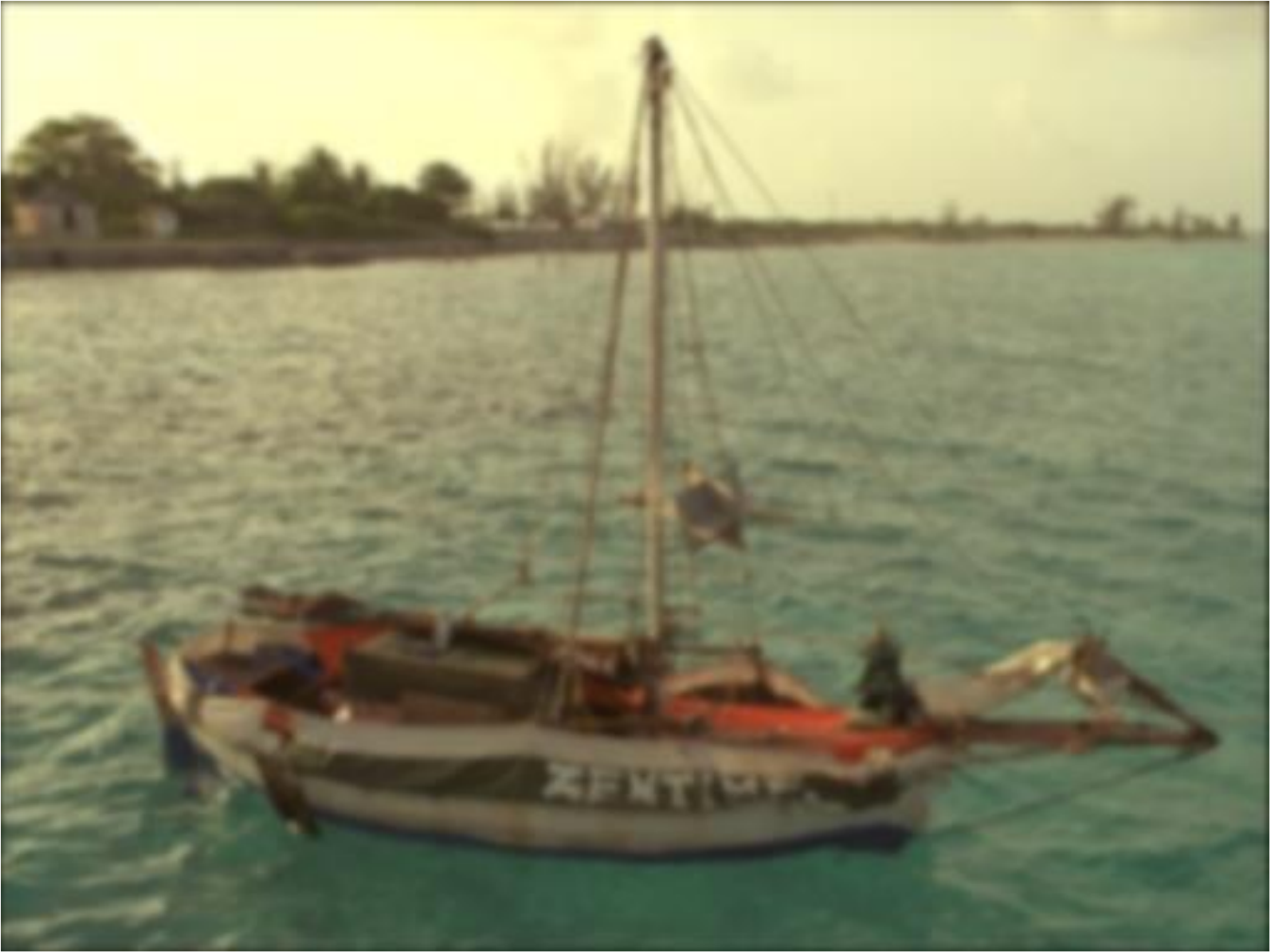}}\hskip.2em
    \subfloat[]{\includegraphics[width=0.32\textwidth]{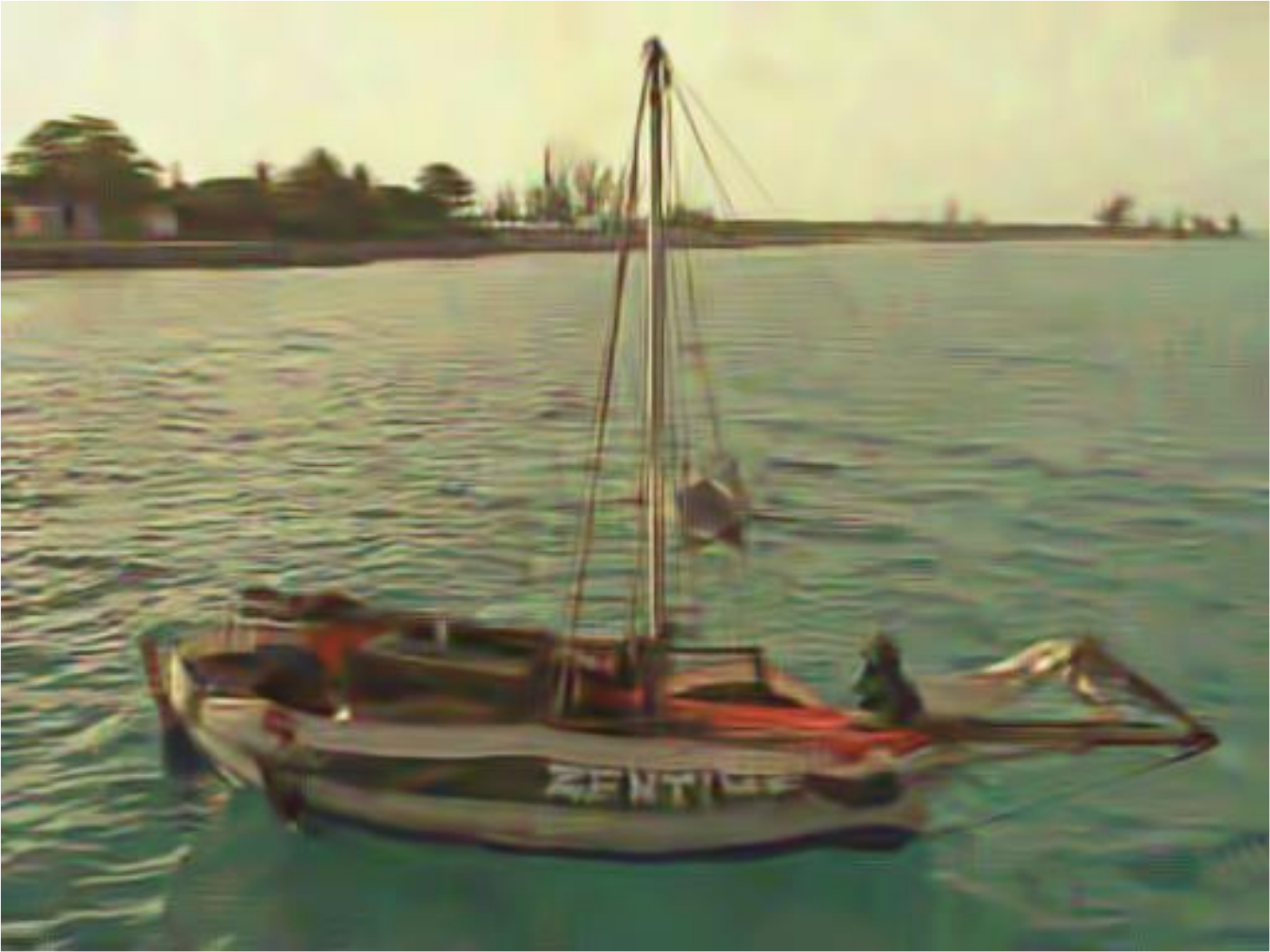}}\hskip.2em
    \subfloat[]{\includegraphics[width=0.32\textwidth]{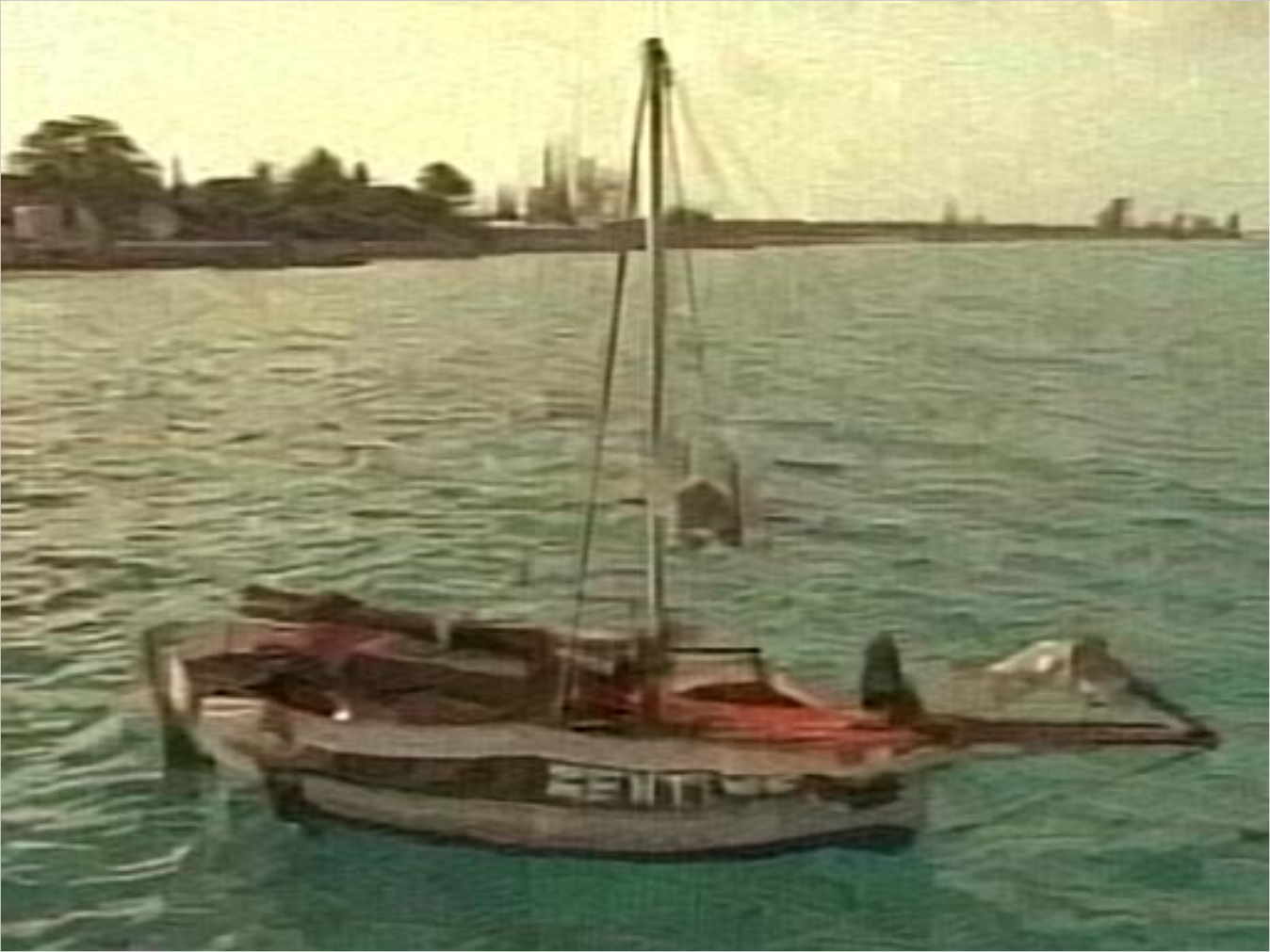}}
     \vspace{2pt}
    \subfloat[]{\includegraphics[width=0.32\textwidth]{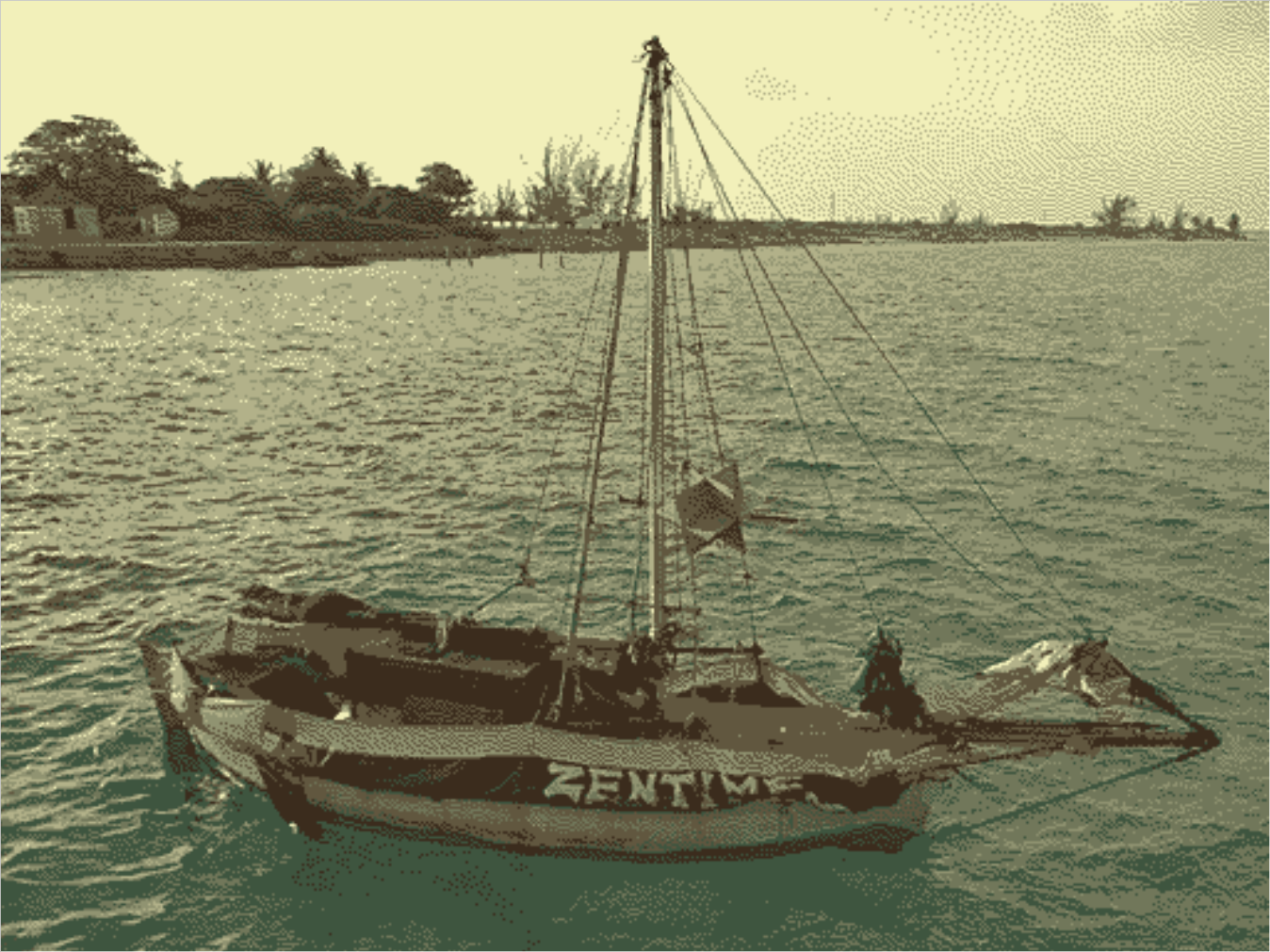}}\hskip.2em
    \subfloat[]{\includegraphics[width=0.32\textwidth]{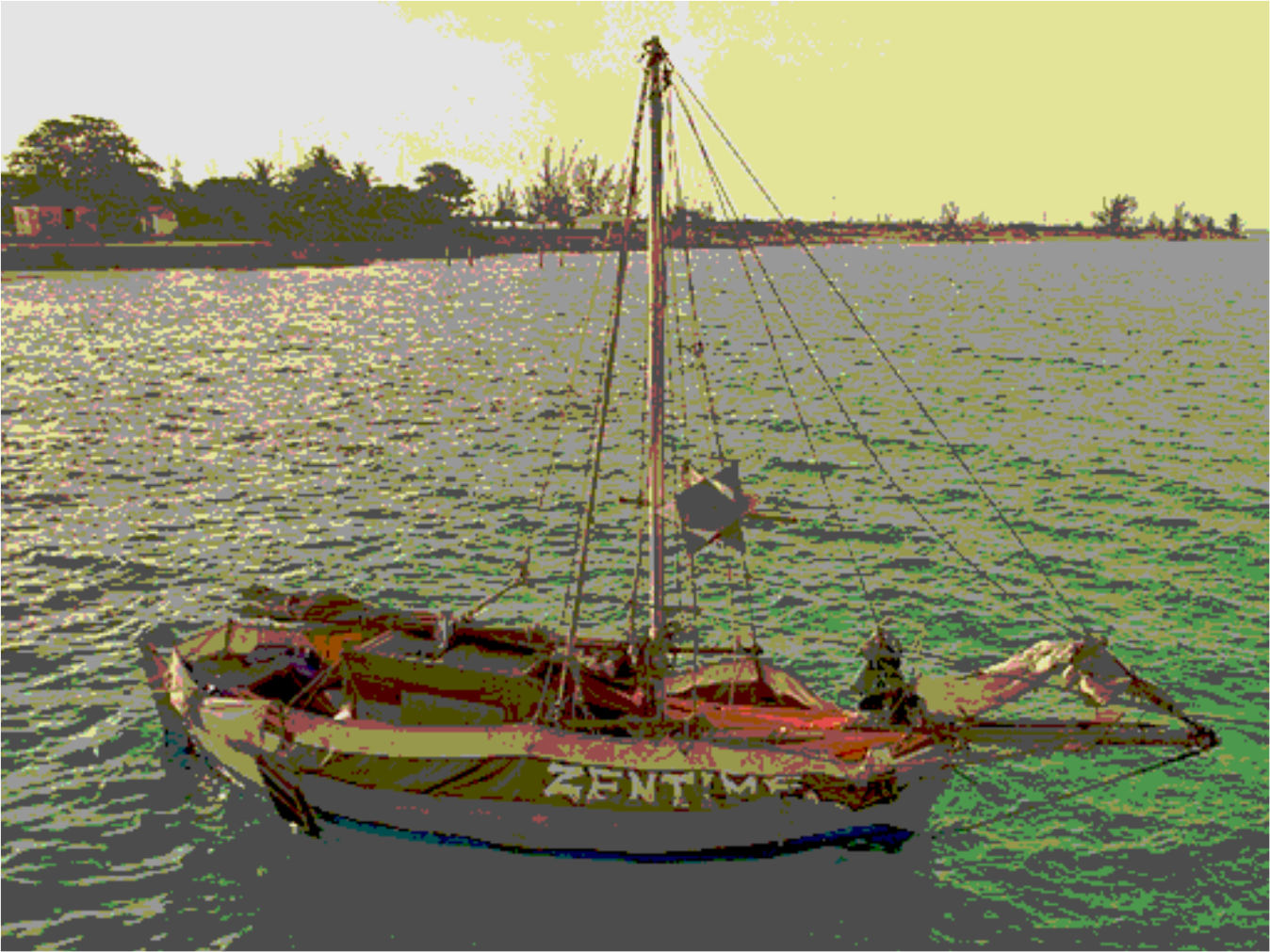}}

    \caption{Images with different distortion types may share similar visual appearances. (a) Additive Gaussian noise. (b) Additive noise in color components. (c) High frequency noise. (d) Gaussian blur. (e) Image denoising. (f) Sparse sampling and reconstruction. (g) Image color quantization with dither. (h) Quantization noise.}\label{fig:tid}
\end{figure*}

\begin{table*}[t]
  \centering
  \caption{SRCC results in a cross-database setting}\label{tab:cross_validation}
  \begin{tabular}{l|ccc|ccc}
      \toprule
        Training & \multicolumn{3}{c|}{LIVE~\cite{sheikh2006statistical}} & \multicolumn{3}{c}{CSIQ~\cite{larson2010most}}\\
     \hline
            Testing & CSIQ & TID2013 & LIVE Challenge & LIVE & TID2013 & LIVE Challenge \\
     \hline
            BRISQUE~\cite{mittal2012no} & 0.562 & 0.358 & 0.337 & 0.847 & 0.454 & 0.131  \\
      M3~\cite{xue2014blind}& 0.621 & 0.344 & 0.226 & 0.797 & 0.328 & 0.183 \\
      FRIQUEE~\cite{ghadiyaram2017perceptual} & {\bf 0.722} & 0.461 & 0.411 & {\bf 0.879} & {\bf 0.463} & 0.264 \\
      CORNIA~\cite{ye2012unsupervised} & 0.649 & 0.360 & 0.443 & 0.853 & 0.312 & {\bf 0.393} \\
      HOSA~\cite{xu2016blind} & 0.594 & 0.361 & {\bf 0.463} & 0.773 & 0.329 & 0.291 \\
            DIQaM~\cite{bosse2016deep} & 0.681 & 0.392 & --- & --- & --- & --- \\
            WaDIQaM~\cite{bosse2016deep} & 0.704 & {\bf 0.462} & --- & --- & --- & --- \\
     \hline
            DB-CNN& {\bf 0.758} & {\bf 0.524} & {\bf 0.567} & {\bf 0.877} & {\bf 0.540} & {\bf 0.452} \\
     \midrule
            Training & \multicolumn{3}{c|}{TID2013~\cite{ponomarenko2013color}} & \multicolumn{3}{c}{LIVE Challenge~\cite{ghadiyaram2016massive}}\\
     \hline
            Testing & LIVE  & CSIQ & LIVE Challenge & LIVE & CSIQ & TID2013 \\
     \hline
            BRISQUE~\cite{mittal2012no} & 0.790 & 0.590 & 0.254 & 0.238 & 0.241 & 0.280  \\
      M3~\cite{xue2014blind}& {\bf 0.873} & 0.605 & 0.112 & 0.059 & 0.109 & 0.058 \\
      FRIQUEE~\cite{ghadiyaram2017perceptual} & 0.755 & 0.635 & 0.181 & {\bf 0.644} & {\bf 0.592} & {\bf 0.424} \\
      CORNIA~\cite{ye2012unsupervised} & 0.846 & 0.672 & 0.293 & 0.588 & 0.446 & 0.403 \\
      HOSA~\cite{xu2016blind} & 0.846 & 0.612 & {\bf 0.319} & 0.537 & 0.336 & 0.399 \\
            DIQaM~\cite{bosse2016deep} & --- & 0.717 & --- & --- & --- & --- \\
            WaDIQaM~\cite{bosse2016deep} & --- & {\bf 0.733} & --- & --- & --- & --- \\
     \hline
            DB-CNN & {\bf 0.891} & {\bf 0.807} & {\bf 0.457} & {\bf 0.746} & {\bf 0.697} & {\bf 0.424} \\
     \bottomrule
   \end{tabular}
\end{table*}

\subsubsection{Performance on Individual Distortion Types}\label{subsec:individual_distortions}
To take a closer look at the behaviors of DB-CNN on individual distortion types along with several competing BIQA models, we test them on a specific distortion type and show the results on LIVE~\cite{sheikh2006statistical}, CSIQ~\cite{larson2010most}, and TID2013~\cite{ponomarenko2013color} in Tables~\ref{tab:live_idd},~\ref{tab:csiq_idd}, and~\ref{tab:tid_idd}, respectively. We find that DB-CNN is among the top two performing models $34$  out of $46$ times. Specifically, on CSIQ, DB-CNN outperforms other counterparts by a large margin, especially for pink noise and contrast change, validating the effectiveness of pre-training in DB-CNN. Although we do not synthesize as many distortion types as in TID2013, we find that DB-CNN performs well on unseen distortion types that exhibit similar artifacts in our pre-training set. As shown in Fig.~\ref{fig:tid}, grainy noise exists in images distorted by additive Gaussian noise, additive noise in color components, and high frequency noise; Gaussian blur, image denoising, and sparse sampling and reconstruction mainly introduce blur; image color quantization with dither and quantization noise also share similar appearances. Trained on synthesized images with additive Gaussian noise, Gaussian blur, and image color quantization with dither, DB-CNN  generalizes well to unseen distortions with similar perceived artifacts. In addition, all BIQA models fail in three distortion types on TID2013, \textit{i.e.}, non-eccentricity pattern noise, local block-wise distortions, and mean shift, whose characteristics are difficult to model.

\subsubsection{Performance across Different Databases}\label{subsec:cross_database}
In this subsection, we evaluate DB-CNN in a cross-database setting against knowledge-driven and CNN-based models. We train knowledge-driven models on one database and test them on the other databases. The results of CNN-based counterparts are reported if available from the original papers.
We show the SRCC results in Table~\ref{tab:cross_validation}, where we see that models trained on LIVE are much easier to generalize to CSIQ and vice versa than other cross-database pairs. When trained on TID2013 and tested on the other two synthetic databases, DB-CNN significantly outperforms the rest  models. However, it is evident that models trained on synthetic databases do not generalize to the authentic LIVE Challenge Database. Despite this, DB-CNN still achieves higher prediction accuracies under such a challenging experimental setup.

\subsubsection{Results on the Waterloo Exploration Database~\cite{ma2017waterloo}}\label{subsec:waterlooeval}
Although SRCC and PLCC have been widely used as the performance criteria in IQA research, they cannot be applied to arbitrarily large databases due to the absence of the ground truths. Three testing criteria are introduced along with the Waterloo Exploration Database in~\cite{ma2017waterloo},~\textit{i.e.}, the pristine/distorted image discriminability test (D-Test), the listwise ranking consistency test (L-Test), and the pairwise preference consistency test (P-Test). D-Test measures the capability of BIQA models in discriminating distorted images from pristine ones. L-Test measures the listwise ranking consistency of BIQA models when rating images with the same content and distortion type but different degradation levels. P-Test measures the pairwise concordance of BIQA models on image pairs with clearly discriminable perceptual quality. More details of the three criteria can be found in~\cite{ma2017waterloo}. Here we use them to test the robustness of DB-CNN on the Waterloo Exploration Database. To ensure the independence of image content during training and testing, we re-train the S-CNN stream in DB-CNN using the distorted images generated from the PASCAL VOC Database only. Experimental results are tabulated in Table~\ref{tab:waterloo}, where we observe that DB-CNN is competitive in all the three tests.

\begin{table}[t]
  \centering
  \caption{Results on the Waterloo Exploration Database~\cite{ma2017waterloo}}\label{tab:waterloo}
  \begin{tabular}{l|cccc}
      \toprule
        Model & D-Test & L-Test & P-Test \\
     \hline
            BRISQUE~\cite{mittal2012no} & 0.9204 & {\bf 0.9772} & 0.9930 \\
            M3~\cite{xue2014blind} & 0.9203 & 0.9106 & 0.9748 \\
      CORNIA~\cite{ye2012unsupervised}& 0.9290 & 0.9764 & 0.9947 \\
      HOSA~\cite{xu2016blind} & 0.9175 & 0.9647 & 0.9983 \\
           dipIQ~\cite{ma2017dipiq} & 0.9346 & {\bf 0.9846} & {\bf 0.9999} \\
            deepIQA~\cite{bosse2016deep} & 0.9074 & 0.9467 & 0.9628 \\

            MEON~\cite{Ma2018End} & {\bf 0.9384} & 0.9669 & 0.9984 \\
     \hline
            DB-CNN & {\bf 0.9616} & 0.9614 & {\bf 0.9992} \\
     \bottomrule
   \end{tabular}
\end{table}

\begin{figure*}
  \centering
  \includegraphics[width=1\textwidth]{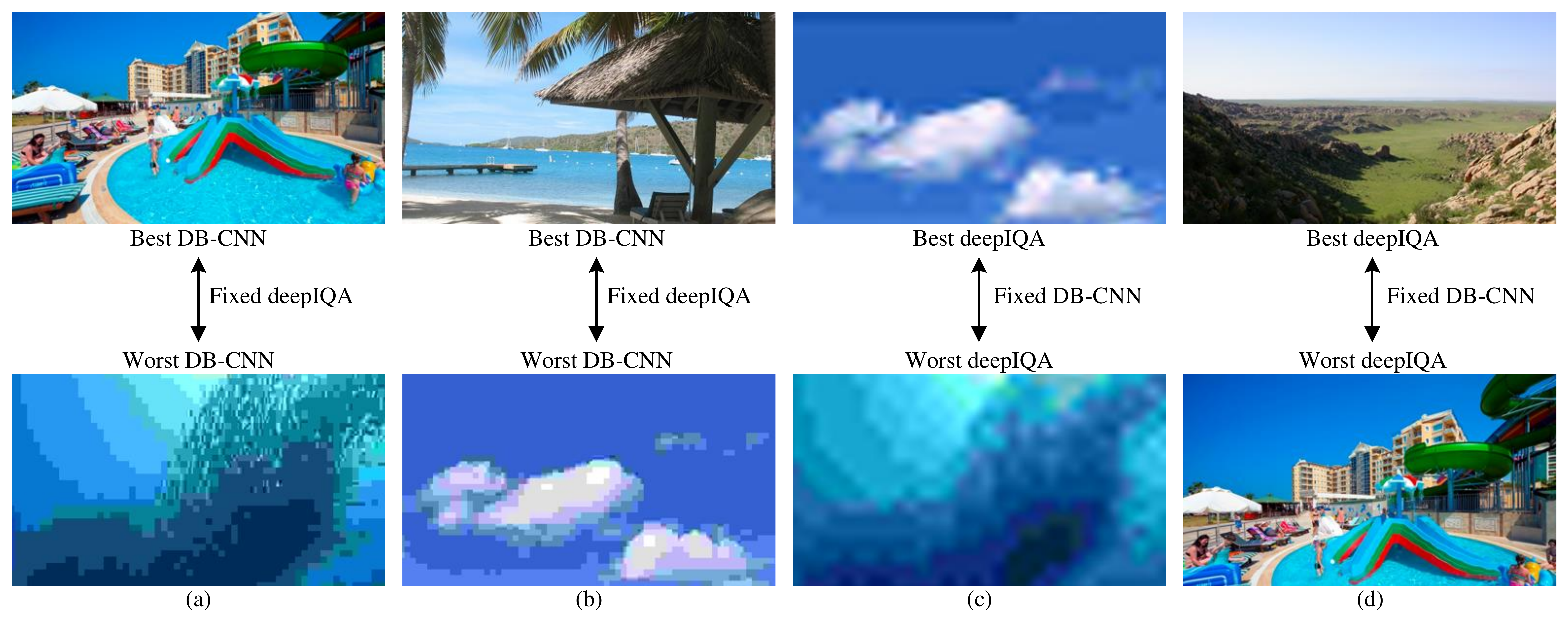}
  \caption{gMAD competition results between DB-CNN and deepIQA~\cite{bosse2016deep}. (a) Fixed deepIQA at the low-quality level. (b) Fixed deepIQA at the high-quality level. (c) Fixed DB-CNN at the low-quality level. (d) Fixed DB-CNN at the high-quality level.}\label{fig:gmad1}
\end{figure*}

\begin{figure*}
  \centering
  \includegraphics[width=1\textwidth]{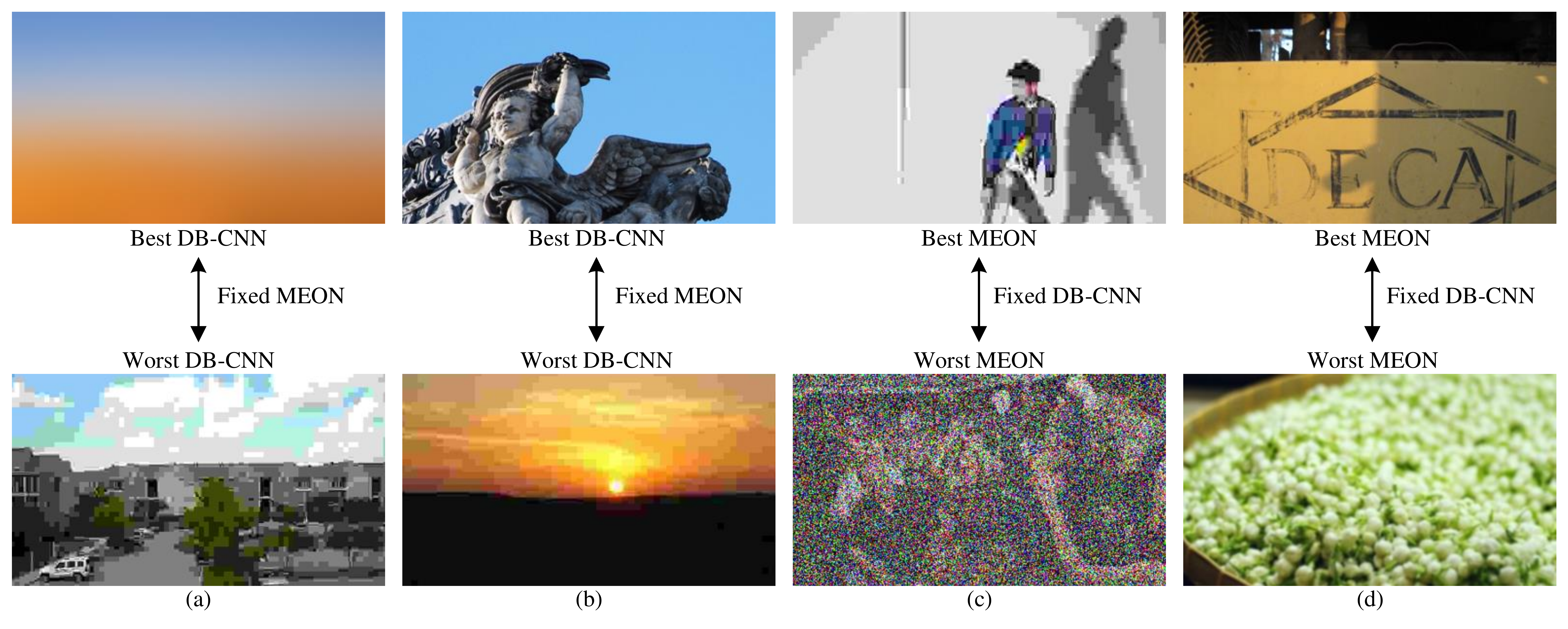}
  \caption{gMAD competition results between DB-CNN and MEON~\cite{Ma2018End}. (a) Fixed MEON at the low-quality level. (b) Fixed MEON at the high-quality level. (c) Fixed DB-CNN at the low-quality level. (d) Fixed DB-CNN at the high-quality level.}\label{fig:gmad2}
\end{figure*}

We further let CNN-based BIQA models play the gMAD competition game~\cite{ma2016group} on the Waterloo Exploration Database~\cite{ma2017waterloo}.
gMAD extends the idea of the MAD competition~\cite{wang2008maximum} and allows a group of IQA models to be falsified in the most efficient way by letting them compete on a large-scale database with no human annotations. A small number of extremal image pairs are  generated automatically by maximizing the responses of the attacker model while fixing the defender model. In Fig.~\ref{fig:gmad1}, DB-CNN first plays the attacker role and deepIQA~\cite{bosse2016deep} acts as the defender. deepIQA~\cite{bosse2016deep} considers pairs (a) and (b) to have the same perceptual quality at the low- and high-quality level, respectively, which is in disagreement with human perception. By contrast, DB-CNN correctly predicts the better quality of  the top images in pairs (a) and (b). We then switch the roles of DB-CNN and deepIQA to obtain pairs (c) and (d). deepIQA fails to falsify DB-CNN, where the two images in one extremal image pair indeed exhibit similar quality. Furthermore, we let DB-CNN fight against MEON~\cite{Ma2018End} and show four extremal image pairs in Fig.~\ref{fig:gmad2}. From pairs (a) and (c), we find that  both DB-CNN and MEON successfully defend the attack from the other model at the low-quality level.
 As for the high-quality level, DB-CNN shows slightly advantage by finding the counterexample of MEON in pair (b). This reveals that MEON does not handle JPEG compression well enough, especially when the image contains few structures. MEON also finds a counterexample of DB-CNN in pair (d), where the bottom image is blurrier than the top one. Through gMAD, there is no clear winner between DB-CNN and MEON, but we identify the weaknesses of the two models.

\subsubsection{Ablation Experiments}\label{subsubsec:ablation}
In order to evaluate the design rationality of DB-CNN, we conduct several ablation experiments with the setups and protocols following Section~\ref{subsec:expsetup}. We first work with a baseline version, where only one stream (either S-CNN or VGG-16) is included. The bilinear pooling is kept, which turns out to be the outer-product of the activations from the last convolution layer with themselves. We then replace the bilinear pooling module with a simple feature concatenation and ensure that the last fully connected layer has approximately the same parameters as in DB-CNN. From Table~\ref{tab:ablation}, we observe that S-CNN and VGG-16 can only deliver promising performance on synthetic and authentic databases, respectively. By contrast, DB-CNN is capable of handling both synthetic and authentic distortions. We also train two DB-CNN models, one from scratch and the other using the distortion type information only during pre-training S-CNN, to validate the necessity of the pre-training stages. From the table, we observe that
with perceptually more meaningful initializations, DB-CNN achieves better performance.

\begin{table}[t]
  \centering
  \caption{Average SRCC results of ablation experiments across ten sessions. ``Scratch'' means DB-CNN is trained from scratch with random initializations. ``distype'' means the S-CNN stream is pre-trained to classify distortion types only, ignoring the distortion level information}\label{tab:ablation}
  \begin{tabular}{l|cccc}
      \toprule
        \multirow{2}{*}{SRCC} & LIVE & CSIQ & TID2013 & LIVE \\
        &\cite{sheikh2006statistical} &~\cite{larson2010most}&\cite{ponomarenko2013color}&Challenge~\cite{ghadiyaram2016massive}\\
     \hline
        S-CNN & \bf{0.963} & {\bf 0.950} & {\bf0.810} & 0.680 \\
        VGG-16 & 0.943 & 0.824 & 0.758 & {\bf0.848} \\
    \hline
     Concatenation &  0.951 & 0.856 & 0.701 & 0.811 \\
     \hline
     DB-CNN scratch &  0.875 & 0.541 & 0.488 & 0.625 \\

     \hline
    DB-CNN distype&  {\bf0.963} & 0.928 & 0.761 & --- \\
     \hline
    DB-CNN& {\bf 0.968} & {\bf0.946} & {\bf 0.816} & {\bf 0.851} \\
     \bottomrule
   \end{tabular}
\end{table}
\section{Conclusion}\label{sec:conclution}
We propose a CNN-based BIQA model for both synthetic and authentic distortions by conceptually modeling them as two-factor variations. DB-CNN demonstrates superior performance, which we believe arises from the two-stream architecture for distortion modeling, pre-training for better initializations, and bilinear pooling for feature combination. Through the validations across different databases, the experiments on the Waterloo Exploration Database, and the results from the gMAD competition, we have shown the scalability, generalizability, and robustness of the proposed DB-CNN model.

DB-CNN is versatile and  extensible. For example, more distortion types and levels can be added to the pre-training set; more sophisticated designs of S-CNN and more powerful CNNs such as ResNet~\cite{he2016deep} can be utilized. One may also improve DB-CNN by considering other variants of bilinear pooling~\cite{gao2016compact}.

The current work deals with synthetic and authentic distortions separately by fine-tuning DB-CNN on either synthetic or authentic databases. How to extend DB-CNN towards a more unified BIQA model, especially in the early feature extraction stage, is an interesting direction yet to be explored.

\bibliographystyle{IEEEtran}
\bibliography{Weixia}

\begin{thebibliography}{10}
\providecommand{\url}[1]{#1}
\csname url@samestyle\endcsname
\providecommand{\newblock}{\relax}
\providecommand{\bibinfo}[2]{#2}
\providecommand{\BIBentrySTDinterwordspacing}{\spaceskip=0pt\relax}
\providecommand{\BIBentryALTinterwordstretchfactor}{4}
\providecommand{\BIBentryALTinterwordspacing}{\spaceskip=\fontdimen2\font plus
\BIBentryALTinterwordstretchfactor\fontdimen3\font minus
  \fontdimen4\font\relax}
\providecommand{\BIBforeignlanguage}[2]{{%
\expandafter\ifx\csname l@#1\endcsname\relax
\typeout{** WARNING: IEEEtran.bst: No hyphenation pattern has been}%
\typeout{** loaded for the language `#1'. Using the pattern for}%
\typeout{** the default language instead.}%
\else
\language=\csname l@#1\endcsname
\fi
#2}}
\providecommand{\BIBdecl}{\relax}
\BIBdecl

\bibitem{bovik2010handbook}
A.~C. Bovik, \emph{Handbook of Image and Video Processing}.\hskip 1em plus
  0.5em minus 0.4em\relax Academic Press, 2010.

\bibitem{BalleLS16a}
\BIBentryALTinterwordspacing
J.~Ball{\'{e}}, V.~Laparra, and E.~P. Simoncelli, ``End-to-end optimized image
  compression,'' \emph{CoRR}, vol. abs/1611.01704, 2016. [Online]. Available:
  \url{http://arxiv.org/abs/1611.01704}
\BIBentrySTDinterwordspacing

\bibitem{duanmu2017quality}
Z.~Duanmu, K.~Ma, and Z.~Wang, ``Quality-of-experience of adaptive video
  streaming: Exploring the space of adaptations,'' in \emph{ACM Multimedia},
  2017, pp. 1752--1760.

\bibitem{wang2006modern}
Z.~Wang and A.~C. Bovik, \emph{Modern Image Quality Assessment}.\hskip 1em plus
  0.5em minus 0.4em\relax Morgan \& Claypool, 2006.

\bibitem{rehman2015display}
A.~Rehman, K.~Zeng, and Z.~Wang, ``Display device-adapted video
  quality-of-experience assessment,'' in \emph{Human Vision and Electronic
  Imaging}, 2015, pp. 1--11.

\bibitem{wang2011reduced}
Z.~Wang and A.~C. Bovik, ``Reduced-and no-reference image quality assessment:
  The natural scene statistic model approach,'' \emph{IEEE Signal Processing
  Magazine}, vol.~28, no.~6, pp. 29--40, Nov. 2011.

\bibitem{mittal2012no}
A.~Mittal, A.~K. Moorthy, and A.~C. Bovik, ``No-reference image quality
  assessment in the spatial domain,'' \emph{IEEE Transactions on Image
  Processing}, vol.~21, no.~12, pp. 4695--4708, Dec. 2012.

\bibitem{ye2012unsupervised}
P.~Ye, J.~Kumar, L.~Kang, and D.~Doermann, ``Unsupervised feature learning
  framework for no-reference image quality assessment,'' in \emph{IEEE
  Conference on Computer Vision and Pattern Recognition}, 2012, pp. 1098--1105.

\bibitem{kang2014convolutional}
L.~Kang, P.~Ye, Y.~Li, and D.~Doermann, ``Convolutional neural networks for
  no-reference image quality assessment,'' in \emph{IEEE Conference on Computer
  Vision and Pattern Recognition}, 2014, pp. 1733--1740.

\bibitem{Ma2018End}
K.~Ma, W.~Liu, K.~Zhang, Z.~Duanmu, Z.~Wang, and W.~Zuo, ``End-to-end blind
  image quality assessment using deep neural networks,'' \emph{IEEE
  Transactions on Image Processing}, vol.~27, no.~3, pp. 1202--1213, Mar. 2018.

\bibitem{deng2009imagenet}
J.~Deng, W.~Dong, R.~Socher, L.-J. Li, K.~Li, and F.-F. Li, ``{ImageNet}: A
  large-scale hierarchical image database,'' in \emph{IEEE Conference on
  Computer Vision and Pattern Recognition}, 2009, pp. 248--255.

\bibitem{kim2017deep}
J.~Kim, H.~Zeng, D.~Ghadiyaram, S.~Lee, L.~Zhang, and A.~C. Bovik, ``Deep
  convolutional neural models for picture-quality prediction: Challenges and
  solutions to data-driven image quality assessment,'' \emph{IEEE Signal
  Processing Magazine}, vol.~34, no.~6, pp. 130--141, Nov. 2017.

\bibitem{ghadiyaram2016massive}
D.~Ghadiyaram and A.~C. Bovik, ``Massive online crowdsourced study of
  subjective and objective picture quality,'' \emph{IEEE Transactions on Image
  Processing}, vol.~25, no.~1, pp. 372--387, Jan. 2016.

\bibitem{sheikh2006statistical}
H.~R. Sheikh, M.~F. Sabir, and A.~C. Bovik, ``A statistical evaluation of
  recent full reference image quality assessment algorithms,'' \emph{IEEE
  Transactions on Image Processing}, vol.~15, no.~11, pp. 3440--3451, Nov.
  2006.

\bibitem{ponomarenko2013color}
N.~Ponomarenko, L.~Jin, O.~Ieremeiev, V.~Lukin, K.~Egiazarian, J.~Astola,
  B.~Vozel, K.~Chehdi, M.~Carli, F.~Battisti, and C.-C.~J. Kuo, ``Image
  database {TID2013}: Peculiarities, results and perspectives,'' \emph{Signal
  Processing: Image Communication}, vol.~30, pp. 57--77, Jan. 2015.

\bibitem{kim2017fully}
J.~Kim and S.~Lee, ``Fully deep blind image quality predictor,'' \emph{IEEE
  Journal of Selected Topics in Signal Processing}, vol.~11, no.~1, pp.
  206--220, Feb. 2017.

\bibitem{Kang2015Simultaneous}
L.~Kang, P.~Ye, Y.~Li, and D.~Doermann, ``Simultaneous estimation of image
  quality and distortion via multi-task convolutional neural networks,'' in
  \emph{IEEE International Conference on Image Processing}, 2015, pp.
  2791--2795.

\bibitem{ma2017waterloo}
K.~Ma, Z.~Duanmu, Q.~Wu, Z.~Wang, H.~Yong, H.~Li, and L.~Zhang, ``Waterloo
  {E}xploration {D}atabase: New challenges for image quality assessment
  models,'' \emph{IEEE Transactions on Image Processing}, vol.~26, no.~2, pp.
  1004--1016, Feb. 2017.

\bibitem{everingham2010pascal}
M.~Everingham, L.~Van~Gool, C.~K. Williams, J.~Winn, and A.~Zisserman, ``The
  {P}ascal {V}isual {O}bject {C}lasses ({VOC}) {C}hallenge,''
  \emph{International Journal of Computer Vision}, vol.~88, no.~2, pp.
  303--338, Jun. 2010.

\bibitem{ghadiyaram2017perceptual}
D.~Ghadiyaram and A.~C. Bovik, ``Perceptual quality prediction on authentically
  distorted images using a bag of features approach,'' \emph{Journal of
  Vision}, vol.~17, no.~1, pp. 32--32, Jan. 2017.

\bibitem{simonyan2014very}
K.~Simonyan and A.~Zisserman, ``Very deep convolutional networks for
  large-scale image recognition,'' in \emph{International Conference on
  Learning Representations}, 2015.

\bibitem{lin2015bilinear}
T.-Y. Lin, A.~RoyChowdhury, and S.~Maji, ``Bilinear {CNN} models for
  fine-grained visual recognition,'' in \emph{IEEE International Conference on
  Computer Vision}, 2015, pp. 1449--1457.

\bibitem{ma2016group}
K.~Ma, Q.~Wu, Z.~Wang, Z.~Duanmu, H.~Yong, H.~Li, and L.~Zhang, ``Group {MAD}
  competition $-$ a new methodology to compare objective image quality
  models,'' in \emph{IEEE Conference on Computer Vision and Pattern
  Recognition}, 2016, pp. 1664--1673.

\bibitem{bosse2016deep}
S.~Bosse, D.~Maniry, K.~R. M¨¹ller, T.~Wiegand, and W.~Samek, ``Deep neural
  networks for no-reference and full-reference image quality assessment,''
  \emph{IEEE Transactions on Image Processing}, vol.~27, no.~1, pp. 206--219,
  Jan. 2018.

\bibitem{ma2017dipiq}
K.~Ma, W.~Liu, T.~Liu, Z.~Wang, and D.~Tao, ``dip{IQ}: Blind image quality
  assessment by learning-to-rank discriminable image pairs,'' \emph{IEEE
  Transactions on Image Processing}, vol.~26, no.~8, pp. 3951--3964, Aug. 2017.

\bibitem{tang2014blind}
H.~Tang, N.~Joshi, and A.~Kapoor, ``Blind image quality assessment using
  semi-supervised rectifier networks,'' in \emph{IEEE Conference on Computer
  Vision and Pattern Recognition}, 2014, pp. 2877--2884.

\bibitem{Bianco2016on}
S.~Bianco, L.~Celona, P.~Napoletano, and R.~Schettini, ``On the use of deep
  learning for blind image quality assessment,'' \emph{CoRR}, vol.
  abs/1602.05531, 2016.

\bibitem{zhang2011fsim}
L.~Zhang, L.~Zhang, X.~Mou, and D.~Zhang, ``F{SIM}: A feature similarity index
  for image quality assessment,'' \emph{IEEE Transactions on Image Processing},
  vol.~20, no.~8, pp. 2378--2386, Aug. 2011.

\bibitem{Ma2015Perceptual}
K.~Ma, K.~Zeng, and Z.~Wang, ``Perceptual quality assessment for multi-exposure
  image fusion,'' \emph{IEEE Transactions on Image Processing}, vol.~24,
  no.~11, pp. 3345--3356, Nov. 2015.

\bibitem{tenenbaum1997separating}
J.~B. Tenenbaum and W.~T. Freeman, ``Separating style and content,'' in
  \emph{Advances in Neural Information Processing Systems}, 1997, pp. 662--668.

\bibitem{simonyan2014two}
K.~Simonyan and A.~Zisserman, ``Two-stream convolutional networks for action
  recognition in videos,'' in \emph{Advances in Neural Information Processing
  Systems}, 2014, pp. 568--576.

\bibitem{fukui2016multimodal}
A.~Fukui, D.~H. Park, D.~Yang, A.~Rohrbach, T.~Darrell, and M.~Rohrbach,
  ``Multimodal compact bilinear pooling for visual question answering and
  visual grounding,'' \emph{CoRR}, vol. abs/1606.01847, 2016.

\bibitem{he2016deep}
K.~He, X.~Zhang, S.~Ren, and J.~Sun, ``Deep residual learning for image
  recognition,'' in \emph{IEEE Conference on Computer Vision and Pattern
  Recognition}, 2016, pp. 770--778.

\bibitem{pennec2006riemannian}
X.~Pennec, P.~Fillard, and N.~Ayache, ``A {R}iemannian framework for tensor
  computing,'' \emph{International Journal of Computer Vision}, vol.~66, no.~1,
  pp. 41--66, Jan. 2006.

\bibitem{larson2010most}
E.~C. Larson and D.~M. Chandler, ``Most apparent distortion: {F}ull-reference
  image quality assessment and the role of strategy,'' \emph{Journal of
  Electronic Imaging}, vol.~19, no.~1, pp. 1--21, Jan. 2010.

\bibitem{Jayaraman2013Objective}
D.~Jayaraman, A.~Mittal, A.~K. Moorthy, and A.~C. Bovik, ``Objective quality
  assessment of multiply distorted images,'' in \emph{Signals, Systems and
  Computers}, 2013, pp. 1693--1697.

\bibitem{video2003final}
\BIBentryALTinterwordspacing
VQEG, ``{Final} report from the video quality experts group on the validation
  of objective models of video quality assessment,'' 2000. [Online]. Available:
  \url{http://www.vqeg.org}
\BIBentrySTDinterwordspacing

\bibitem{he2015delving}
K.~He, X.~Zhang, S.~Ren, and J.~Sun, ``Delving deep into rectifiers: Surpassing
  human-level performance on {I}mage{N}et classification,'' in \emph{IEEE
  International Conference on Computer Vision}, 2015, pp. 1026--1034.

\bibitem{Kingma2014adam}
D.~P. Kingma and J.~Ba, ``Adam: {A} method for stochastic optimization,''
  \emph{CoRR}, vol. abs/1412.6980, 2014.

\bibitem{ioffe2015batch}
S.~Ioffe and C.~Szegedy, ``Batch normalization: Accelerating deep network
  training by reducing internal covariate shift,'' in \emph{International
  Conference on Machine Learning}, 2015, pp. 448--456.

\bibitem{vedaldi2015matconvnet}
A.~Vedaldi and K.~Lenc, ``Mat{C}onv{N}et: Convolutional neural networks for
  {M}atlab,'' in \emph{ACM International Conference on Multimedia}, 2015, pp.
  689--692.

\bibitem{xue2014blind}
W.~Xue, X.~Mou, L.~Zhang, A.~C. Bovik, and X.~Feng, ``Blind image quality
  assessment using joint statistics of gradient magnitude and {L}aplacian
  features,'' \emph{IEEE Transactions on Image Processing}, vol.~23, no.~11,
  pp. 4850--4862, Nov. 2014.

\bibitem{xu2016blind}
J.~Xu, P.~Ye, Q.~Li, H.~Du, Y.~Liu, and D.~Doermann, ``Blind image quality
  assessment based on high order statistics aggregation,'' \emph{IEEE
  Transactions on Image Processing}, vol.~25, no.~9, pp. 4444--4457, Sep. 2016.

\bibitem{wang2008maximum}
Z.~Wang and E.~P. Simoncelli, ``Maximum differentiation ({MAD}) competition: A
  methodology for comparing computational models of perceptual quantities,''
  \emph{Journal of Vision}, vol.~8, no.~12, pp. 8.1--8.13, Sep. 2008.

\bibitem{gao2016compact}
Y.~Gao, O.~Beijbom, N.~Zhang, and T.~Darrell, ``Compact bilinear pooling,'' in
  \emph{IEEE Conference on Computer Vision and Pattern Recognition}, 2016, pp.
  317--326.

\end{thebibliography}

\begin{IEEEbiography}[{\includegraphics[width=1in,height=1.25in,clip,keepaspectratio]{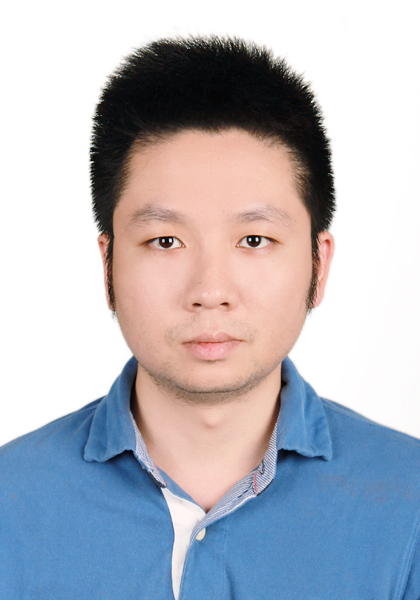}}]{Weixia Zhang}
received the B.E. degree from the Wuhan University, Wuhan, China, in 2011 and the M.S. degree in electrical and computer engineering from the University of Rochester, NY, USA, in 2013. He was a Project Engineer at Cyberspace Great Wall Inc., Beijing, China, in 2014. He then received the Ph.D. degree from the Wuhan University, Wuhan, China, in 2018. He is currently a Postdoctoral Fellow with the Institute of Artificial Intelligence, Shanghai Jiao Tong University. His research interests include image and video quality/aesthetics assessment, image recognition, and vision and language understanding.
\end{IEEEbiography}

\begin{IEEEbiography}[{\includegraphics[width=1in,height=1.25in,clip,keepaspectratio]{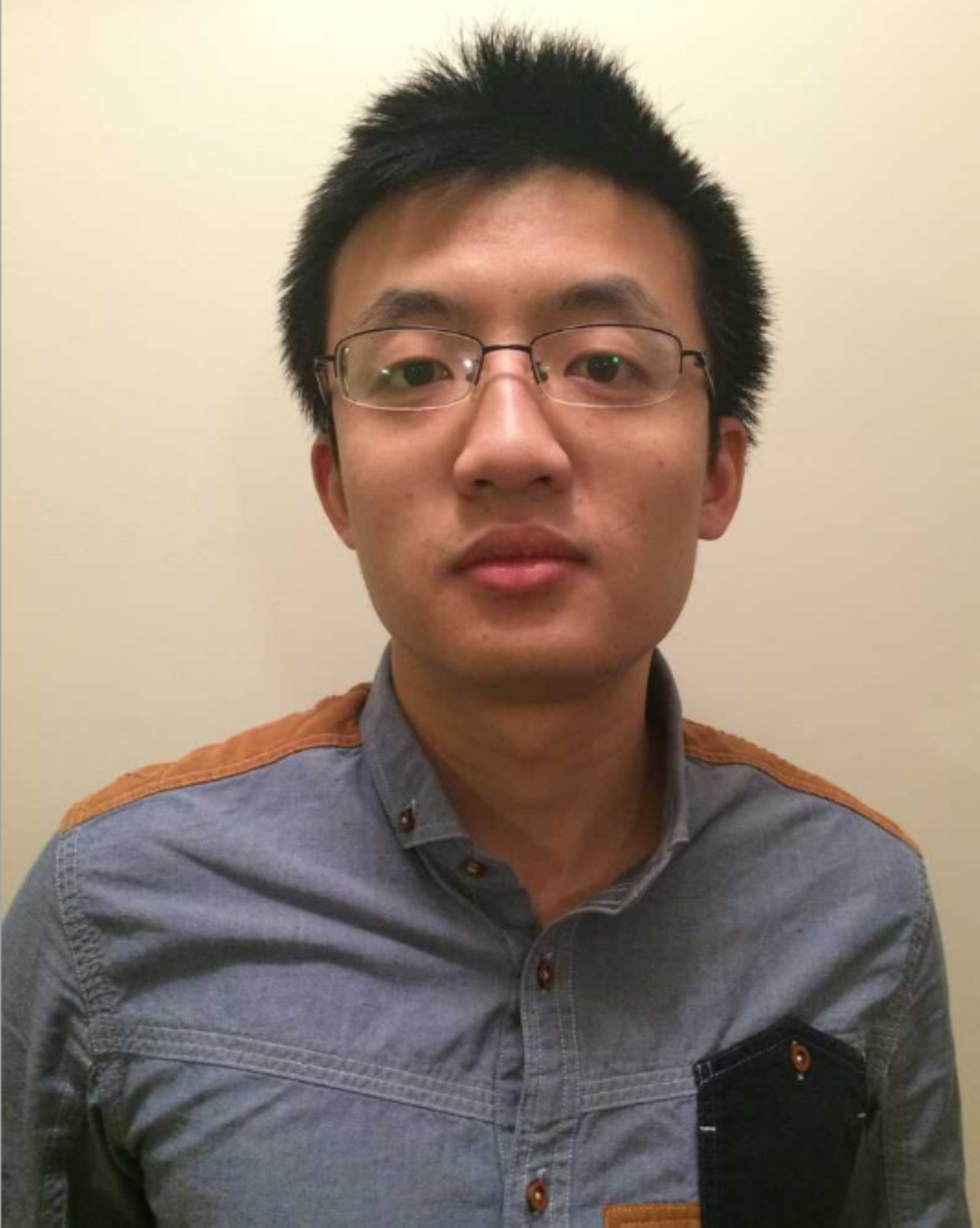}}]{Kede Ma}
(S'13-M'18) received the B.E. degree from the University of Science and Technology of China, Hefei, China, in 2012, and the M.S. and Ph.D. degrees in electrical and computer engineering from the University of Waterloo, Waterloo, ON, Canada, in 2014 and 2017, respectively. He is currently a Research Associate with the Howard Hughes Medical Institute and Laboratory for Computational Vision, New York University, New York, NY, USA. His research interests include perceptual image processing, computational vision, and computational photography.
\end{IEEEbiography}

\begin{IEEEbiography}[{\includegraphics[width=1in,height=1.25in,clip,keepaspectratio]{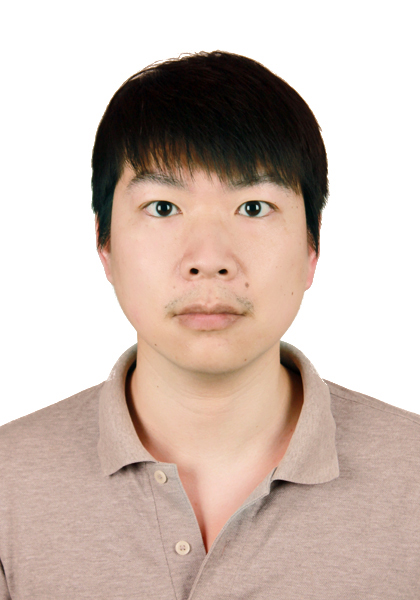}}]{Jia Yan}
was born in Hubei, China. He received the B.E. degree in electronic information science and technology and the Ph.D. degree in communication and information system from the Wuhan University, Wuhan, China, in 2005 and 2010, respectively.
From 2011 to  2014, he was a Postdoctoral Research Fellow with the Center for Physics of the Earth Geophysics, Wuhan University, Wuhan, China, where he is an Associate Professor. Dr. Yan is currently a Visiting Scholar with Department of Electrical and Computer Engineering, University of Waterloo, ON, Canada. He has authored or co-authored more than 20 publications in top journals and conference proceedings. He serves as a reviewer of {\em Neurocomputing} and {\em Journal of Visual Communication and Image Representation}. His research interests include blind image quality assessment and image enhancement.
\end{IEEEbiography}

\begin{IEEEbiography}[{\includegraphics[width=1in,height=1.25in,clip,keepaspectratio]{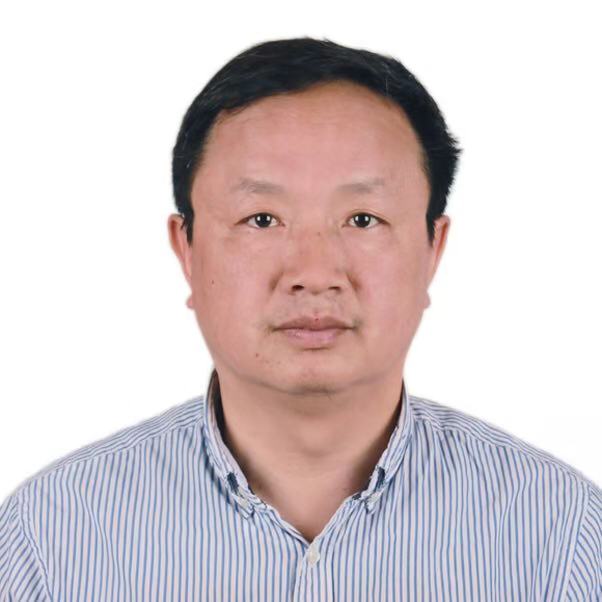}}]{Dexiang Deng}
received the B.E and M.S. degrees from the Wuhan Technical University of Surveying, Wuhan, China, in 1982 and 1985, respectively. He is currently a Professor and a Ph.D. advisor with the Department of Electrical Engineering, Wuhan University, Wuhan, China. His research interests include space image processing, machine vision, and system on chip.
\end{IEEEbiography}

\begin{IEEEbiography}[{\includegraphics[width=1in,height=1.25in,clip,keepaspectratio]{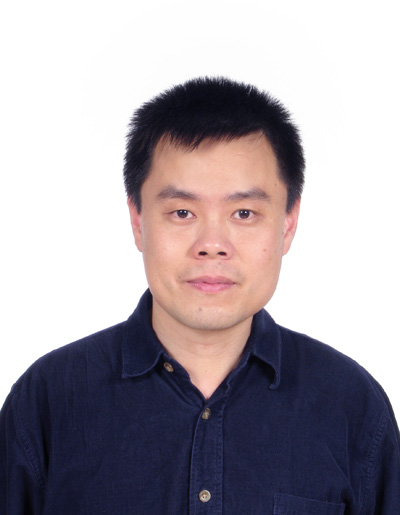}}]{Zhou Wang}
(S'99-M'02-SM'12-F'14) received the Ph.D. degree from The University of Texas at Austin in 2001. He is currently a Professor and University Research Chair in the Department of Electrical and Computer Engineering, University of Waterloo, Canada. His research interests include image and video processing and coding; visual quality assessment and optimization; computational vision and pattern analysis; multimedia communications; and biomedical signal processing. He has more than 200 publications in these fields with over 40,000 citations (Google Scholar).

Dr. Wang serves as a Senior Area Editor of {\em IEEE Transactions on Image Processing} (2015-present). Previously, he served as a member of IEEE Multimedia Signal Processing Technical Committee (2013-2015), an Associate Editor of {\em IEEE Transactions on Circuits and Systems for Video Technology} (2016-2018), an Associate Editor of {\em IEEE Transactions on Image Processing} (2009-2014), {\em Pattern Recognition} (2006-present) and {\em IEEE Signal Processing Letters} (2006-2010), and a Guest Editor of {\em IEEE Journal of Selected Topics in Signal Processing} (2013-2014 and 2007-2009). He is a Fellow of Royal Society of Canada and Canadian Academy of Engineering, and a recipient of 2017 Faculty of Engineering Research Excellence Award at University of Waterloo, 2016 IEEE Signal Processing Society Sustained Impact Paper Award, 2015 Primetime Engineering Emmy Award, 2014 NSERC E.W.R. Steacie Memorial Fellowship Award, 2013 IEEE Signal Processing Magazine Best Paper Award, and 2009 IEEE Signal Processing Society Best Paper Award.
\end{IEEEbiography}
\end{document}